\documentclass[11pt]{article}
\usepackage{amsmath,amssymb,amsfonts}
\usepackage[english]{babel}
\makeatletter
\@addtoreset{equation}{section}
\makeatother

\def\nn{\nonumber} \def\bd{\begin{document}} \def\ed{\end{document}}
\def\ds{\documentstyle}
\let\bm=\bibitem
\newcommand{\be}{\begin{equation}}
\newcommand{\ee}{\end{equation}}
\newcommand{\bea}{\setlength\arraycolsep{2pt} \begin{eqnarray}}
\newcommand{\eea}{\end{eqnarray}}
\newcommand{\hoch}[1]{$\, ^{#1}$}
\def\p{\partial}

\title{\large {\bf Conserved charge of a gravity theory with $p$-form gauge
fields and its property under Kaluza-Klein reduction}}
\date{}

\author{Jun-Jin Peng\footnote{pengjjph@163.com}  \\ \\
\small \sl School of Physics and Electronic Science, Guizhou Normal University,\\
\small Guiyang, Guizhou 550001, People's Republic of China
}

\begin{document}
\maketitle
\vspace{20pt}

\begin{abstract}
In this paper, we investigate the conserved charges of generally
diffeomorphism invariant gravity theories with a wide variety of matter fields,
particularly of the theories with multiple scalar fields and $p$-form potentials,
in the context of the off-shell generalized Abbott-Deser-Tekin (ADT) formalism.
We first construct a new off-shell ADT current that consists
of the terms for the variation of a Killing vector and expressions of the field
equations as well as the Lie derivative of a surface term with respect to the
Killing vector within the framework of generally diffeomorphism invariant gravity
theories involving various matter fields. After deriving the off-shell ADT
potential corresponding to this current, we propose a formula of conserved
charges for these theories. Next, we derive the off-shell
ADT potential associated with the generic Lagrangian that describes a large range
of gravity theories with a number of scalar fields and $p$-form potentials. Finally,
the properties of the off-shell generalized ADT charges for the theory of Einstein
gravity and the gravity theories with a single $p$-form potential
are investigated by performing Kaluza-Klein dimensional
reduction along a compactified direction. The results indicate that the charge
contributed by all the fields in the lower-dimensional theory is equal to that of the
higher-dimensional one at mathematical level with the hypothesis that the
higher-dimensional spacetime allows for the existence of the compactified dimension.
In order to illustrate our calculations, the mass and angular momentum for the
five-dimensional rotating Kaluza-Klein black holes are explicitly evaluated as an
example.
\end{abstract}

\newpage
\tableofcontents
\voffset=-.90pt
\vspace{40pt}


\section{Introduction}

The definition of conserved charges in various gravity theories is of great
importance for the understanding of many physics properties of spacetime. As
is well known, its remarkable and successful
applications are those in spacetime thermodynamics and black hole physics. Due to
this, much work has been devoted to seeking feasible approaches to define the
conserved charges of gravity theories appropriately. A rather effective route to
do this is to make use of Noether theorem.

Till now several approaches in terms of the Noether procedure have been proposed to
compute the conserved charges. One of them is the Abbott-Deser-Tekin (ADT) method
\cite{AbbottD,DeserT,DeserT2}. The ADT formalism, which is defined in terms of the Noether
potential got through the linearized perturbation for the expression of equation
of motion in a fixed background of (A)dS spacetime, has made some progress on
computation scheme for conserved charges of asymptotically (A)dS black holes in various
gravity theories. Since the background metric is a vacuum solution of the field
equation, the Noether potential in the original ADT formalism is on-shell.
One of the applications for the ADT formalism is to compute the conserved charges
within the context of three-dimensional topologically massive gravity in the
work \cite{DeserTmassG}, which was subsequently extended to nonasymptotically AdS black
holes in the same thoery in \cite{BouchClem}. Notably, in the latter, although the current
is on-shell, the potential was extracted from the expression of the current derived
without the requirement that the background metric is on-shell. This form of the current
allows for the possibility of the construction of an off-shell current in the context of
pure gravity theories.
Afterwards, the development for the derivation of the potential was extended
to compute the mass and angular momentum of three-dimensional Chern-Simons
black holes \cite{MouClGuL,MouCGuCSBH} and black holes in three-dimensional new
massive gravity \cite{NamParkY}.

Recently, in Ref. \cite{KimKY}, relieving the constraint that the background metric
satisfies the field equations, Kim, Kulkarni and Yi proposed a quasi-local formulation of
conserved charges within the framework of generic covariant pure gravity theories by
constructing an off-shell ADT current to generalize the conventional on-shell Noether
potential in the original ADT formalism to off-shell level, as well as following
works \cite{BarnichB,Barnich,BarnichC} to incorporate a single
parameter path in the space of solutions into their definition. These modifications
make it more operable to evaluate the Noether potential in terms of the corresponding current
and the procedure of computation become more convenient to manipulate. Owing to these,
the off-shell generalized formalism for the quasi-local conserved charges provides another
fruitful way to evaluate the ADT charges for various theories of gravity and it has been
extended to investigate the conserved charges of a wide scope of gravity theories with or without
matter fields
\cite{CiteHJPY,JJPeng,SetAdam,JJPXcai,JJPIJMpA,LiLuWei,Wuli,JJPengPLB,NamPark3D,CiteHPY,SetaAdChernS,CiteChernS}.
Particularly, in \cite{CiteHJPY}, the method in \cite{KimKY} was generalized
to the gravity theories in the presence of matter fields.

In the work \cite{JJPeng}, reconsidering the conserved charges of generally diffeomorphism
invariant gravity theories with only gravitational field, we constructed an off-shell
generalized ADT current that takes a different form from the one given in \cite{KimKY} through
the linear combination of the variation of the Bianchi identity for the expression of the
field equation and the Lie derivative for the variation of the Lagrangian along a Killing
vector. However, both the currents are equivalent since the Lie derivative of a surface
term with respect to the Killing vector disappears. Our procedure to construct the off-shell
ADT current is rather simple and the current naturally yields its corresponding potential
that completely coincides with the one in \cite{KimKY}. A question naturally arises, i.e.
whether the procedure in \cite{JJPeng} is applicable to construct the off-shell ADT current
and potential of diffeomorphism invariant gravity theories in the presence of a wide variety
of matter fields.

On the other hand, $p$-form gauge fields appear systematically within the context of
the string theories, the higher-dimensional supergravity theories as well as the brane
theories, for example, in the ten-dimensional type IIB and eleven-dimensional supergravity
theories. They have attracted much attention in recent years. In particular, some exact
solutions in the supergravity theories have been found. For several of them see the references
\cite{ChowComp,Wu3CBH,TomMizog,GodelBHWuGi,ChongCLP,WuKKsdsBH}. To understand the behaviour
of the $p$-form gauge fields on the conserved charges for gravity theories with such
fields, it is of great interest to apply the off-shell generalized ADT formalism to
fully identify their contribution. If we do this, the outcome will be able to provide
a basis for further investigation on thermodynamics and other related properties of the 
solutions found in these theories. Besides, as is well-known, if the
higher-dimensional spacetime allows for the existence of a compactified dimension, 
by performing Kaluza-Klein dimensional reduction along this
compactified direction, the higher-dimensional gravity theories can be described by
reduced gravity theories involving a series of lower-dimensional fields, such as the
gravitational field, the Kaluza-Klein vector, the dilaton field and so on. Through such
a procedure, a natural question is how the conserved charges associated with the
higher-dimensional theories of gravity to behave in the lower-dimensional reduced theories
after the dimensional reduction. We wonder wether a consistent Kaluza-Klein dimensional
reduction can yield consistent conserved charges of the gravity theories.

In consideration of the above issues, the main motivation of the present work is to shed
some light on the conserved charges of
generally diffeomorphism invariant gravity theories with various matter fields, particularly
of the theories with a number of scalar fields and $p$-form potentials, in the context
of the off-shell generalized ADT formalism proposed in \cite{KimKY,CiteHJPY}. To achieve
this, along the line of the work \cite{JJPeng}, the off-shell ADT formalism is extended
to these gravity theories by constructing a new generic off-shell ADT current without the
condition that the Killing vector associated with the conserved charge has to be fixed.
In terms of the off-shell ADT potential constructed from this current, the formula for
the conserved charges is presented. In particular, we propose a very general Lagrangian
that describes a large range of gravity theories with scalar fields and $p$-form
potentials. Accordingly, the conserved charges for this Lagrangian are explicitly
investigated. What is more, another motivation is to gain a good understanding of
the properties for the conserved charges under the Kaluza-Klein dimensional reduction.
This is achieved by exploiting the off-shell ADT charges of $(D+1)$-dimensional pure
Einstein gravity and
gravity theories with a single $p$-form gauge field before and after performing Kaluza-Klein
reduction on an $S^1$ circle. The results demonstrate that the off-shell ADT charges
exhibit the property of invariance under the dimensional reduction.

The remainder of this paper is organized as follows: In section \ref{secone}, we
derive the general formalism for conserved charges of diffeomorphism covariant
gravity theories with arbitrary matter fields by constructing the off-shell
generalized ADT currents and potentials for such theories in a new manner.
In section \ref{sectwo}, we move on to apply the general formalism of conserved
charges in the previous section to generic covariant gravity theories in the
presence of multiple scalar fields and $p$-form potentials. In section
\ref{secthree}, we study the
properties of conserved charges of $(D+1)$-dimensional Einstein gravity
by performing Kaluza-Klein reduction on a circle. In section \ref{secfour},
the analysis on Einstein gravity is extended to investigate the behaviour
of conserved charges within the framework of gravity theories
with multiple scalar fields and a single $(n-1)$-form gauge field. In order to
illustrate our calculations, we consider to explicitly compute the conserved
charges of five-dimensional rotating Kaluza-Klein black holes in light of the
off-shell generalized ADT formalism within section \ref{secfive}.
The last section is our conclusions and discussions on future work. To gain
more details on our derivations in the main body of the text, several appendixes
are presented. Particularly,
in Appendix \textbf{\ref{appendA}}, the off-shell Noether currents and potentials for
the Lagrangians with multiple scalar fields and $p$-form potentials are calculated
in details. The Appendix\textbf{ \ref{appendD}} is devoted to the discussion of Kaluza-Klein
dimensional reduction to the generic Lagrangian that consists of a number of scalar
fields, multiple $p$-form potentials, as well as a gravitational field, along a
compactified direction.

\section{The off-shell generalized ADT currents and conserved charges of
diffeomorphism invariant gravity theories with arbitrary matter fields}\label{secone}
In our previous work \cite{JJPeng}, providing a new route varying the Bianchi identity
for the expression of the field equations, together with help of the Killing equation for a
Killing vector that generates a spacetime symmetry, we got another
form for the off-shell generalized ADT current of generally
diffeomorphism covariant gravity theories with only gravitational field given in
\cite{KimKY}. The new form of the current contains an additional term that is just
one half of the Lie derivative of a surface term with respect to the Killing
vector. However, in essence, this current makes no difference to the one
presented in \cite{KimKY} because of the vanishing of the Lie derivative
for the surface term.

In this section, we generalize the derivation in the work \cite{JJPeng}
to gravity theories with a wide variety of matter fields, satisfying the
criteria of general covariance. As a result, we propose a generalized
off-shell ADT current that arises from the diffeomorphism symmetry of
the Lagrangian and the vanishing of the divergence for the contraction
between the Killing vector and a generally non-symmetric $2$-rank tensor,
which is always made up of expressions of the equations of motion for
gravitational field and matter fields. The newly improved current,
including the Lie derivative of the surface term with respect to the Killing
vector and extra terms involving the variation of the Killing vector in
comparison with the one given in \cite{JJPeng}, differs from the
current proposed in \cite{CiteHJPY}, where the off-shell generalized ADT
formalism for the pure gravity theories in the work \cite{KimKY} was
extended to study conserved charges of covariant gravity theories in the
presence of matter fields. Moving on to construct out the off-shell ADT
potential from the current, we further present a formula for the conserved
charges of these gravity theories along the line of the work \cite{KimKY}.

We proceed by considering the Lagrangian describing $D$-dimensional generally
diffeomorphism covariant gravity theories with various matter fields
\be
\mathcal{L}= \sqrt{-g}L\big(g_{\mu\nu},\psi^{(r)}\big)
\, , \label{GenLagran}
\ee
in which the quantity $\psi^{(r)}$ is taken to run over all the matter fields of the
theories. The variation of the general Lagrangian (\ref{GenLagran}) with respect to
the gravitational field and all the matter fields reads
\be
\delta\mathcal{L}=\sqrt{-g}\Big(\mathcal{E}^{(gr)}_{\mu\nu}\delta g^{\mu\nu}
+\sum_r \mathcal{E}_{\psi^{(r)}}\delta\psi^{(r)}
+\nabla_\mu \Theta^\mu\big(\delta g,\delta \psi^{(r)}\big)\Big)
\, , \label{VarGenLagran}
\ee
where $\mathcal{E}^{(gr)}_{\mu\nu}$ and $\mathcal{E}_{\psi^{(r)}}$
are the expressions for the field equations, and
$\Theta^\mu\big(\delta g,\delta \psi^{(r)}\big)$ is a surface term.
When the variation of the gravitation field $g_{\mu\nu}$ and the
matter fields $\psi^{(r)}$ behaves as their Lie derivative along
any smooth vector field $\zeta^\mu$, namely,
$\delta g_{\mu\nu}\rightarrow \mathcal{L}_\zeta g_{\mu\nu}$
and $\delta\psi^{(r)}\rightarrow \mathcal{L}_\zeta \psi^{(r)}$
under diffeomorphisms $x^\mu\rightarrow x^\mu-\zeta^\mu$
generated by the vector field $\zeta^\mu$,
it was shown in \cite{CiteHJPY,IWaldentro} that the following
identity
\be
2\nabla_\mu (\mathcal{E}^{\mu\nu}\zeta_{\nu})
=\mathcal{E}_{(gr)}^{\mu\nu}\mathcal{L}_\zeta g_{\mu\nu}
-\sum_r \mathcal{E}_{\psi^{(r)}}\mathcal{L}_\zeta\psi^{(r)}
\label{generalizedBianch}
\ee
generally holds for diffeomorphism invariant gravity theories. The identity
(\ref{generalizedBianch}) plays a key role for deriving the off-shell Noether
current. Its left-hand side is the divergence for
the contraction of the vector $\zeta_\nu$ with the $2$-rank tensor
$\mathcal{E}^{\mu\nu}$, which is in general unnecessary to be symmetric
and is defined by
\be
\mathcal{E}^{\mu\nu}=\mathcal{E}_{(gr)}^{\mu\nu}
-\frac{1}{2}Z^{\mu\nu}_{(\psi)}
\, ,\label{Emunudefin}
\ee
where the $2$-rank tensor $Z^{\mu\nu}_{(\psi)}$ is generally proportional to
some combination of the expressions of the field equations $\mathcal{E}_{\psi^{(r)}}$
and it vanishes when $\mathcal{E}_{\psi^{(r)}}=0$. Particularly,
in the absence of matter fields, Eq. (\ref{generalizedBianch}) yields
$\nabla_\mu\mathcal{E}_{(gr)}^{\mu\nu}=0$, which is just the Bianchi
identity for covariant gravity theories with only gravitational field
\cite{JJPeng}. Replacing
the variation in Eq. (\ref{VarGenLagran}) by the Lie derivative along
the vector field $\zeta^\mu$, as well as making use of Eq. (\ref{generalizedBianch}),
we obtain the following conservation law
\be
\nabla_\mu J^\mu=0 \, ,
\ee
where the divergence-free vector $J^\mu$, defined by
\be
J^\mu=2\mathcal{E}^{\mu\nu}\zeta_\nu+\zeta^\mu L
-\Theta^\mu\big(\mathcal{L}_\zeta g,\mathcal{L}_\zeta \psi^{(r)}\big)
\, , \label{Genoffshcurr}
\ee
is an off-shell Noether current \cite{KimKY,CiteHJPY,TPadoffshNOe}
in the sense that the conservation law for this Noether current holds without
the assumption that the equations of motion for the fields are valid.
Furthermore, the off-shell Noether potential $K^{\mu\nu}$ corresponding to
the current can be defined as
\be
J^\mu=\nabla_\nu K^{\mu\nu}
\, . \label{Genoffshpoten}
\ee

Next, we focus our attention on detailed derivation of the off-shell generalized
ADT current and its corresponding potential on basis of the above off-shell Noether
current and potential. In the case where the vector $\zeta^\mu$ is replaced by
a Killing vector $\xi^\mu$ that satisfies the following symmetry conditions:
\bea
\mathcal{L}_\xi g_{\mu\nu}&=&\nabla_\mu \xi_\nu+\nabla_\nu \xi_\mu
=0
\, , \nn \\
\mathcal{L}_\xi \psi^{(r)} &=&0
\, , \label{GenKillvec}
\eea
the right-hand side of Eq. (\ref{generalizedBianch}) vanishes, which directly
gives rise to
\be
\nabla_\mu (\mathcal{E}^{\mu\nu}\xi_{\nu})=0
\, , \label{Emunuxiidenti}
\ee
i.e. the divergence for the contraction of the $\mathcal{E}^{\mu\nu}$
tensor and the Killing vector $\xi_\nu$ takes zero value. This can be regarded as a
consequence of the diffeomorphism and symmetry of matter fields. Significantly,
one will see that the identity (\ref{Emunuxiidenti}) plays a vital role in
constructing the off-shell generalized ADT current in the following. Varying
this identity, we get
\be
\delta[\nabla_\mu (\mathcal{E}^{\mu\nu}\xi_{\nu})]
=\nabla_\mu \mathcal{J}^\mu_{[1]}=0
\, , \label{VarBIone}
\ee
where the current
\be
\mathcal{J}^\mu_{[1]}=
\delta\mathcal{E}^{\mu\nu}\xi_\nu
+\mathcal{E}^{\mu\nu}\xi^\lambda\delta g_{\nu\lambda}
+\frac{1}{2}\mathcal{E}^{\mu\nu}\xi_\nu g^{\rho\sigma}\delta g_{\rho\sigma}
+\mathcal{E}^{\mu}_{~\nu}\delta\xi^\nu
\, . \label{currecalJ1}
\ee
Unlike in \cite{KimKY,CiteHJPY,JJPeng}, here and in what follows we do not
impose the constraint of the vanishing of the variation for the Killing
vector $\xi^\mu$, i.e., $\delta\xi^\mu=0$, on any current, although this
generally holds for the Killing vectors associated with energy and angular
momentum. In particular, when all the fields satisfy the equations of motion,
namely, $\mathcal{E}_{(gr)}^{\mu\nu}=0=\mathcal{E}_{\psi^{(r)}}$, the current
$\mathcal{J}^\mu_{[1]}$ becomes the original on-shell ADT current \cite{AbbottD,DeserT}.

On the other hand, letting the Lie derivative with respect to the Killing vector
$\xi^\mu$ act on the variation equation (\ref{VarGenLagran}) of the Lagrangian,
we have
\bea
\mathcal{L}_\xi\delta(\sqrt{-g}L)
&=&\sqrt{-g}\nabla_\mu\Big[\xi^\mu\mathcal{E}^{(gr)}_{\rho\sigma}\delta g^{\rho\sigma}
+\xi^\mu\sum_r \mathcal{E}_{\psi^{(r)}}\delta\psi^{(r)}
+\mathcal{L}_\xi\Theta^\mu\big(\delta g,\delta \psi^{(r)}\big) \Big]
\nn \\
&=&-\sqrt{-g}\nabla_\mu (L\delta \xi^\mu)
\, , \label{LiedeVarGenLan}
\eea
which yields a conserved current that reads
\be
\mathcal{J}^\mu_{[2]}=
\xi^\mu\mathcal{E}^{(gr)}_{\rho\sigma}\delta g^{\rho\sigma}
+\xi^\mu\sum_r \mathcal{E}_{\psi^{(r)}}\delta\psi^{(r)}
+\mathcal{L}_\xi\Theta^\mu\big(\delta g,\delta \psi^{(r)}\big)
+L\delta \xi^\mu
\, . \label{CurreLievarLag}
\ee
This current, together with the $\mathcal{J}^\mu_{[1]}$ current in
Eq. (\ref{currecalJ1}), should be able to underlie the construction of
a new Noether current. In fact, through a linear combination of
the currents $\mathcal{J}^\mu_{[1]}$ and $\mathcal{J}^\mu_{[2]}$,
a new off-shell generalized ADT current can be proposed as
\bea
\mathcal{J}_{ADT}^\mu&=&
\mathcal{J}^\mu_{[1]}+\frac{1}{2}\mathcal{J}^\mu_{[2]}
\nn \\
&=&\delta\mathcal{E}^{\mu\nu}\xi_\nu
+\mathcal{E}^{\mu\nu}\xi^\lambda\delta g_{\nu\lambda}
+\frac{1}{2}\mathcal{E}^{\mu\nu}\xi_\nu g^{\rho\sigma}\delta g_{\rho\sigma}
+\frac{1}{2}\xi^\mu\mathcal{E}^{(gr)}_{\rho\sigma}\delta g^{\rho\sigma}
\nn \\
&&+\frac{1}{2}\xi^\mu\sum_r \mathcal{E}_{\psi^{(r)}}\delta\psi^{(r)}
+\frac{1}{2}\mathcal{L}_\xi\Theta^\mu\big(\delta g,\delta \psi^{(r)}\big)
\nn \\
&&+\frac{1}{2}L\delta \xi^\mu
+\mathcal{E}^{\mu}_{~\nu}\delta\xi^\nu
 \, , \label{OffShADTCurr}
\eea
where the current $\mathcal{J}_{ADT}^\mu$ can be thought of as a generalization
of the off-shell ADT current for pure gravity theories in \cite{JJPeng} to
the one for covariant gravity theories with arbitrary matter fields when
the variation of the Killing vector $\delta\xi^\mu$ vanishes, namely,
$\delta\xi^\mu=0$. In addition, in comparison with the off-shell ADT current
for the gravity theories with matter fields in \cite{CiteHJPY}, the current
in Eq. (\ref{OffShADTCurr}), containing the term
$\mathcal{L}_\xi\Theta^\mu\big(\delta g,\delta \psi^{(r)}\big)$ as well as
the terms with the variation of the Killing vector, gives rise to a
different form from the one given in \cite{CiteHJPY}. However,  both of the
currents are essentially equivalent since one can verify that
$\mathcal{L}_\xi\Theta^\mu\big(\delta g,\delta \psi^{(r)}\big)=0$ for the
generally diffeomorphism covariant Lagrangian (\ref{GenLagran}) under the
condition that the variation of the $\xi^\mu$ Killing vector, obeying
Eq. (\ref{GenKillvec}), disappears.

Our procedure to bring in the off-shell ADT current $\mathcal{J}_{ADT}^\mu$
is very simple since it is only constructed out of a combination for two
generally satisfied equations, that is, the variation equation (\ref{VarBIone})
of a divergence-free term made up of expressions of the field equations and
a Killing vector, together with the equation (\ref{LiedeVarGenLan}) for the
Lie derivative of the variation of the Lagrangian along the Killing vector.
It is also unnecessary for us to fix the Killing
vector in order to get the off-shell ADT current and the latter potential.
What is more, we shall see that our formulation for the off-shell ADT current
in turn makes a natural and practical construction for derivation of its
corresponding potential in the following.

In terms of the off-shell ADT current $\mathcal{J}_{ADT}^\mu$ given by
Eq. (\ref{OffShADTCurr}), we go on to derive its corresponding potential
$\mathcal{Q}_{ADT}^{\mu\nu}$, which is associated with the current
through the well-known relation
\be
\mathcal{J}_{ADT}^\mu=\nabla_\nu \mathcal{Q}_{ADT}^{\mu\nu}
\, . \label{GenOfShNP}
\ee
To do this, the $\mathcal{J}_{ADT}^\mu$ current multiplied by the factor
$\sqrt{-g}$ is expressed as
\bea
\sqrt{-g}\mathcal{J}_{ADT}^\mu
&=&\delta\Big(\sqrt{-g}\mathcal{E}^{\mu\nu}\xi_\nu\Big)
+\frac{1}{2}\sqrt{-g}\xi^\mu\mathcal{E}^{(gr)}_{\rho\sigma}\delta g^{\rho\sigma}
+\frac{1}{2}\sqrt{-g}\xi^\mu\sum_r \mathcal{E}_{\psi^{(r)}}\delta\psi^{(r)}
\nn \\
&&+\frac{1}{2}\sqrt{-g}\big(\mathcal{L}_\xi\Theta^\mu\big(\delta g,\delta \psi^{(r)}\big)
+L\delta\xi^\mu \big)
 \, . \label{GenOffADTcurr}
\eea
According to Eq. (\ref{Genoffshcurr}), the off-shell Noether current
associated with the Killing vector $\xi^\mu$ is given by
$J^\mu_\xi=J^\mu(\zeta\rightarrow \xi)$. Like in \cite{TPadman,JJPeng},
the $J^\mu_\xi$ current can also be defined as the equivalent form,
i.e. $J^\mu_\xi=2\mathcal{E}^{\mu\nu}\xi_\nu+\xi^\mu L$, stemming from
that the disappearance of the Lie derivative for the gravitational field
and all the matter fields with respect to the Killing vector leads to
the vanishing of the
$\Theta^\mu\big(\mathcal{L}_\xi g,\mathcal{L}_\xi \psi^{(r)}\big)$
term for a generally diffeomorphism invariant theory.
Substituting the off-shell Noether current $J^\mu_\xi$
and the Lie derivative of the surface term
$\Theta^\mu\big(\delta g,\delta \psi^{(r)}\big)$ given by
\be
\mathcal{L}_\xi\Theta^\mu\big(\delta g,\delta \psi^{(r)}\big)= -2\nabla_\nu
\Big(\xi^{[\mu}\Theta^{\nu]}\big(\delta g,\delta \psi^{(r)}\big)\Big)
+\xi^\mu\nabla_\nu\Theta^\nu\big(\delta g,\delta \psi^{(r)}\big) \,
\ee
into Eq. (\ref{GenOffADTcurr}), we reexpress the current $\mathcal{J}_{ADT}^\mu$
as the form
\bea
\sqrt{-g}\mathcal{J}_{ADT}^\mu
&=&\frac{1}{2}\delta\Big(\sqrt{-g}J^\mu_\xi\Big)
-\sqrt{-g}\nabla_\nu
\Big(\xi^{[\mu}\Theta^{\nu]}\big(\delta g,\delta \psi^{(r)}\big)\Big)
\nn \\
&&+\frac{1}{2}\delta
\big[\sqrt{-g}\Theta^\mu\big(\mathcal{L}_\xi g,\mathcal{L}_\xi \psi^{(r)}\big)\big]
 \, .\label{GenOffADTcurr2}
\eea
Due to Eq. (\ref{GenOffADTcurr2}), one can observe that the preservation of the
$\mathcal{L}_\xi\Theta^\mu\big(\delta g,\delta \psi^{(r)}\big)$ term makes it rather
natural lead the $\xi^{[\mu}\Theta^{\nu]}\big(\delta g,\delta \psi^{(r)}\big)$ term
into the potential.
Note that the surface term $\Theta^\mu\big(\mathcal{L}_\xi g,\mathcal{L}_\xi \psi^{(r)}\big)=0$
for the generally diffeomorphism covariant Lagrangian (\ref{GenLagran}). We further
send the current $\mathcal{J}_{ADT}^\mu$ into the form
\bea
\sqrt{-g}\mathcal{J}_{ADT}^\mu
&=&\partial_\nu\Big[\frac{1}{2}\delta(\sqrt{-g} K_\xi^{\mu\nu})
-\sqrt{-g}\xi^{[\mu}\Theta^{\nu]}\big(\delta g,\delta \psi^{(r)}\big)\Big]
\nn \\
&=&\partial_\nu\big(\sqrt{-g} \mathcal{Q}_{ADT}^{\mu\nu} \big)
\, , \label{GenJKmunu}
\eea
where $K_\xi^{\mu\nu}= K^{\mu\nu}(\zeta\rightarrow\xi)$ is determined by
Eq. (\ref{Genoffshpoten}). From Eq. (\ref{GenJKmunu}), we draw out the
important relationship between the generalized ADT potential
$\mathcal{Q}_{ADT}^{\mu\nu}$ and the off-shell
Noether potential $K_\xi^{\mu\nu}$, that is,
\be
\sqrt{-g}\mathcal{Q}_{ADT}^{\mu\nu}
=\frac{1}{2}\delta(\sqrt{-g}K_\xi^{\mu\nu})
-\sqrt{-g}\xi^{[\mu}\Theta^{\nu]}\big(\delta g,\delta \psi^{(r)}\big)
 \, . \label{ADTpotenQ}
\ee
Equivalently, one can define the off-shell generalized ADT potential as
follows:
\be
\mathcal{Q}_{ADT}^{\mu\nu}
=\frac{1}{2}\delta K_\xi^{\mu\nu}
+\frac{1}{4}K_\xi^{\mu\nu}g^{\rho\sigma}\delta g_{\rho\sigma}
-\xi^{[\mu}\Theta^{\nu]}\big(\delta g,\delta \psi^{(r)}\big)
\, . \label{ADTpotenQ2}
\ee
Compared with the off-shell ADT potential given in
\cite{KimKY,CiteHJPY,JJPeng}, it is worth noting that the Killing vector
$\xi^{\mu}$ involved in the potential (\ref{ADTpotenQ2})
is unnecessary to be fixed when varying the potential $K_\xi^{\mu\nu}$,
although they take the same forms.
As it was explicitly demonstrated in \cite{CiteHJPY}, the off-shell ADT potential
(\ref{ADTpotenQ2}) is equivalent with the on-shell Noether potential derived
via the well-known covariant phase space approach, proposed by Lee, Iyer and Wald (LIW)
\cite{LeeWald,IyerWald,IWaldentro}, despite the currents are different from
each other. What is more, the ADT potential
is consistent with the one defined via the Barnich-Brandt-Compere (BBC) method
\cite{BarnichB,Barnich,BarnichC,BCintegC}, which developed the covariant phase
space method. Due to these, the off-shell
ADT formalism can be regarded as an equivalent generalization of the LIW method
or the BBC method. Unlike in \cite{KimKY,CiteHJPY}, note that both the LIW and
BBC approaches also do not impose the constraint that the variation of the
Killing vector associated with the potential has to vanish as in the present work.

Furthermore, as it has been shown that the off-shell ADT current (\ref{OffShADTCurr})
arises from the linear combination of the current $\mathcal{J}^\mu_{[1]}$
and one half of the current $\mathcal{J}^\mu_{[2]}$, a new current in terms of
another linear combination of both the two currents can be proposed as
\be
\check{\mathcal{J}}^\mu=
\mathcal{J}^\mu_{[1]}+\frac{1}{2}(2k+1)\mathcal{J}^\mu_{[2]}
\, ,
\ee
where the constant $k$ takes a generic value. A new potential corresponding
to the current $\check{\mathcal{J}}^\mu$ is defined by
\be
\check{\mathcal{Q}}^{\mu\nu}=\mathcal{Q}_{ADT}^{\mu\nu}
+k\mathcal{K}^{\mu\nu}_{[2]}
\, ,\label{PotenchekQmm}
\ee
where the Noether potential $\mathcal{K}^{\mu\nu}_{[2]}$ is determined by the
equation $\mathcal{J}^\mu_{[2]}=\nabla_\nu\mathcal{K}^{\mu\nu}_{[2]}$. According to
Eq. (\ref{PotenchekQmm}), one observes that $\check{\mathcal{Q}}^{\mu\nu}$
returns to the off-shell ADT potential $\mathcal{Q}_{ADT}^{\mu\nu}$ when
$k=0$ and both the potentials $\check{\mathcal{Q}}^{\mu\nu}$ and
$\mathcal{Q}_{ADT}^{\mu\nu}$ coincide with the original on-shell ADT
potential when all the fields satisfy the equations of motion accompanying
with the vanishing of the variation for the Killing vector.

Like in \cite{KimKY}, by following the BBC approach
\cite{BarnichB,Barnich,BarnichC,BCintegC} to incorporate
a single parameter path characterized by a parameter $s$, where $s\in[0,1]$, in the space
of solutions, we define the covariant formulation of conserved charges associated with the
off-shell ADT potential $\mathcal{Q}_{ADT}^{\mu\nu}$ in Eq. (\ref{ADTpotenQ}) by
\be
\mathcal{Q}=\frac{1}{8\pi G_{(D)}}\int_0^1 ds \int d\Sigma_{\mu\nu}
\mathcal{Q}_{ADT}^{\mu\nu}\big(g,\psi^{(r)}; s\big)
\, , \label{QdefineAn}
\ee
where $d\Sigma_{\mu\nu}=\frac{1}{2}\frac{1}{(D-2)!}
\epsilon_{\mu\nu\mu_1\mu_2\cdot\cdot\cdot\mu_{(D-2)}}dx^{\mu_1}\wedge\cdot\cdot\cdot
\wedge dx^{\mu_{(D-2)}}$ with $\epsilon_{012\cdot\cdot\cdot(D-1)}=\sqrt{-g}$ and
$G_{(D)}$ is the gravitational constant in $D$ dimensions.
Equation (\ref{QdefineAn}) can be regarded as a proposal of the formalism for
the conserved charge, defined in the interior region or at the asymptotical infinity,
for any covariant gravity theory with the Lagrangian (\ref{GenLagran}) whenever its
integration is well defined \cite{BCintegC}. In contrast to another formula of
conserved charges for covariant gravity theories in \cite{HajphDth,HajSheik,GhHaSe},
which is defined in terms of the so-called solution phase space method building
on basis of the LIW and BBC approaches, the formula (\ref{QdefineAn}) is equivalent
with that one.

Finally, in contrast with the original ADT formulation \cite{AbbottD,DeserT,DeserT2},
the formula (\ref{QdefineAn}) for the conserved charges is dependent on the gravitational
and matter fields, as well as their fluctuations, while the usual ADT charges merely involve
the gravitational field and its perturbation. Such a manner of the original ADT formulation
may succeed to produce physical conserved
charges when the matter fields fall off quite fast at infinity and they do not change
the default asymptotic structure, which is the usual asymptotically (A)dS space.
Otherwise, a serious consideration on the contributions from the matter fields becomes
necessary to make the ADT formalism more universal, as in the present work. For example,
the original ADT formulation fails to yield the physically meaningful mass of the black holes in
Horndeski theory \cite{JJPengPLB} and the G\"{o}del-type black holes in five-dimensional
minimal supergravity \cite{GodelBHWuGi}. However, the formula (\ref{QdefineAn}) is
applicable since it contains the contributions from the matter fields. What is more,
if the matter fields are taken into account, the ADT formalism naturally possesses the
property that it is invariant under conformal transformation \cite{JJPIJMpA}, but the
usual ADT formalism requires the asymptotic condition that the conformal factor goes
to unity at infinity to have the same property \cite{ConfTofDT}.

\section{Conserved charges of gravity theories with scalar fields and $p$-form
potentials}\label{sectwo}

In this section, our main goal is to extend the general derivations in the previous
section to systematically investigate the conserved charges of $D$-dimensional
covariant gravity theories consisting of a gravitational field $g_{\mu\nu}$,
$m$ scalar fields $\phi^{(k)}$ $(k=1, 2, \cdot\cdot\cdot, m)$ and a number of
$p$-form potentials $A_{(p)}$, where $1\leq p\leq n-1$ and $2\leq n\leq D$.
In \cite{LikopformCC,Rogatpform,Compepform}, the conserved charges of the
gravity theories with $p$-form potentials were also investigated in other methods.
The Lagrangian describing these theories is assumed to take the generic form
\bea
\mathcal{L}&=& \mathcal{L}_R +\mathcal{L}_\phi +\mathcal{L}_M \, , \nn \\
\mathcal{L}_R&=&\sqrt{-g}R \, , \nn \\
\mathcal{L}_\phi&=&\sqrt{-g}L_\phi
=\sqrt{-g}\Big[\sum^m_{i,j=1}X_{ij}\big(\phi^{(k)}\big)
\nabla^\mu\phi^{(i)}\nabla_\mu\phi^{(j)}+V\big(\phi^{(k)}\big)\Big]
\, , \nn \\
\mathcal{L}_M&=&\sqrt{-g}L_M\big(g, \phi^{(k)}, A_{(p)}, F_{(p+1)}\big)
\, , \label{LagranPhikAl}
\eea
where the $(p+1)$-form field strengths are defined by $F_{(p+1)}=dA_{(p)}$
and the pair of the indices $(i,j)$ in the function $X_{ij}$ are symmetric.
In this work,
without loss of generality, the function $L_M$ is supposed to possess two
types of structures. The first one is
\bea
L_M &\sim& W\big(\phi^{(k)}\big)g\cdot\cdot\cdot g
F_{(q_1)}\cdot\cdot\cdot F_{(q_J)}A_{(p_1)}\cdot\cdot\cdot A_{(p_I)}
\, , \nn \\
&&(p_1,\cdot\cdot\cdot,p_I=1,\cdot\cdot\cdot,n-1; q_1,\cdot\cdot\cdot,
q_J=2,\cdot\cdot\cdot,n)
\, , \label{LMform1}
\eea
and the second one is
\bea
L_M &=&  \sqrt{-g}W\big(\phi^{(k)}\big)g^{\mu_1\nu_1}\cdot\cdot\cdot
g^{\mu_D\nu_D}
\bar{\epsilon}_{\mu_1\cdot\cdot\cdot\mu_D}\times
\nn \\
&&\big(F_{(q_1)}\cdot\cdot\cdot F_{(q_t)}A_{(p_1)}\cdot\cdot\cdot A_{(p_s)}
\big)_{\nu_1\cdot\cdot\cdot\nu_D}
\, . \label{LMform2}
\eea
In the above equation, $\bar{\epsilon}_{\mu_1\cdot\cdot\cdot\mu_D}$ is the
totally antisymmetric Levi-Civita tensor density with
$\bar{\epsilon}_{01\cdot\cdot\cdot(D-1)}=1$. The indices
$p_1,\cdot\cdot\cdot,p_s=1,\cdot\cdot\cdot,(n-1)$ and the indices
$q_1,\cdot\cdot\cdot,q_t=2,\cdot\cdot\cdot,n$. Equation (\ref{LMform2}) in general
describes the conventional Chern-Simons-like terms involved in various gauge theories.
In fact, the first structure of $L_M$ in Eq. (\ref{LMform1}) can be generally
rewritten as a more concrete form, which is as follows:
\be
L_M=
\mathbb{L}_M= W\big(\phi^{(k)}\big)H_{(N)}^{\mu_1\cdot\cdot\cdot\mu_N}
Y_{(N)\mu_1\cdot\cdot\cdot\mu_N}
\, , \quad
1\leq N\leq D
\, , \label{LMform1anoth}
\ee
where the two totally antisymmetric $N$-rank tensors $H_{(N)}$ and
$Y_{(N)}$ are defined by
\bea
H_{(N)\mu_1\cdot\cdot\cdot\mu_N}&=&
\big(F_{(q_1)}\cdot\cdot\cdot F_{(q_t)}
A_{(p_1)}\cdot\cdot\cdot A_{(p_s)}\big)_{[\mu_1\cdot\cdot\cdot\mu_N]}
\, , \nn \\
Y_{(N)\mu_1\cdot\cdot\cdot\mu_N}&=&
\big(F_{(\tilde{q}_1)}\cdot\cdot\cdot F_{(\tilde{q}_j)}
A_{(\tilde{p}_1)}\cdot\cdot\cdot A_{(\tilde{p}_i)}\big)_{[\mu_1\cdot\cdot\cdot\mu_N]}
\, . \label{HYNform}
\eea
In the above equation, the integers $\tilde{p}_1,\cdot\cdot\cdot,\tilde{p}_i$
range from $1$ to $(n-1)$ while the integers
$\tilde{q}_1,\cdot\cdot\cdot,\tilde{q}_j$ range from $2$ to $n$. In addition,
all the indices
$(p_1,\cdot\cdot\cdot,p_s,\tilde{p}_1,\cdot\cdot\cdot,\tilde{p}_i)$ and
$(q_1,\cdot\cdot\cdot,q_t,\tilde{q}_1,\cdot\cdot\cdot,\tilde{q}_j)$
have to satisfy the following constraints
\bea
N&=&(p_1+\cdot\cdot\cdot+p_s)+(q_1+\cdot\cdot\cdot+q_t)
\, , \nn \\
N&=&(\tilde{p}_{1}+\cdot\cdot\cdot+\tilde{p}_i)
+(\tilde{q}_{1}+\cdot\cdot\cdot+\tilde{q}_j)
\, .
\eea
Among all the potentials in $H_{(N)}$ and $Y_{(N)}$, some or all of them are
allowed to be equal to each other, as well as all the field strengths.
However, in our analysis, all the potentials and field strengths are
treated as being formally independent of each other. Furthermore, it
is worth noting that the second structure in Eq. (\ref{LMform2}) can
also be reexpressed as the form like the one in Eq. (\ref{LMform1anoth})
if the Levi-Civita tensor density
$\bar{\epsilon}_{\mu_1\cdot\cdot\cdot\mu_D}$ is understood as a
$D$-form ``potential'', but an additional factor $\sqrt{-g}$ appears.
In such a case, one will see that the procedure to derive the expected
results, including the expressions of the equations of motion, currents
and potentials, is parallel with the one for $L_M$ with the
form (\ref{LMform1}) or (\ref{LMform1anoth}) from appendix
\ref{appendA}. To this point, the form (\ref{LMform1anoth}) for $L_M$
is more general.

Let us pause to make a comment on the Lagrangian $\mathcal{L}_M$. Due
to the forms of $L_M$ given in Eqs. (\ref{LMform1}), (\ref{LMform2})
and (\ref{LMform1anoth}), one sees that the Lagrangian $\mathcal{L}_M$ is
very general. It incorporates a broad class of terms consisting of scalar
fields and $p$-form gauge fields for the Lagrangians in the context of
a wide range of gravity theories with these matter fields, such as the
Einstein-Maxwell-dilaton theories, the low-energy effective field theories of
heterotic string theories, supergravity theories and so on. Particularly in
the context of supergravity theories, the parts involving scalar fields and
$p$-form potentials of the Lagrangian, describing the five-dimensional
Einstein-Maxwell-Chern-Simons theory, the bosonic sector of the
ten-dimensional type IIB supergravity, the bosonic part of the
eleven-dimensional supergravity etc can be regarded as special cases of
$\mathcal{L}_M$.

Varying the Lagrangian (\ref{LagranPhikAl}) with respect to the gravitational
field, the scalar fields and the $p$-form potentials, one obtains
\bea
\delta \mathcal{L} &=&\sqrt{-g} \Big[\mathcal{E}^{(gr)}_{\mu\nu}\delta g^{\mu\nu}
+\sum_{k=1}^m \breve{\mathcal{E}}^{(\phi)}_{(k)}\delta\phi^{(k)}
+\sum_{p=1}^{n-1} \mathcal{E}_{(p)}^{(A)\mu_1\cdot\cdot\cdot\mu_p}
\delta A_{(p)\mu_1\cdot\cdot\cdot\mu_p}
\nn \\
&&+\nabla_\mu \Theta^\mu\big(\delta g,\delta\phi^{(k)},\delta A_{(p)}\big)\Big]
\, , \label{VarLagPhikAl}
\eea
where the expressions $\mathcal{E}_{(p)}^{(A)\mu_1\cdot\cdot\cdot\mu_p}$ of the
field equations with respect to the gauge fields $A_{(p)}$ are given
in Eq. (\ref{EOMofLagranLM}), and the other ones for the gravitational field
and scalar fields, as well as the surface term, are defined through
\bea
\mathcal{E}^{(gr)}_{\mu\nu}&=&\mathcal{E}^{(gr)}_{(R)\mu\nu}
+\mathcal{E}^{(gr)}_{(\phi)\mu\nu}+\mathcal{E}^{(gr)}_{(M)\mu\nu}
\, , \label{grEOMofLagPhikAl} \\
\breve{\mathcal{E}}^{(\phi)}_{(k)}&=&\tilde{\mathcal{E}}^{(\phi)}_{(k)}
+\mathcal{E}^{(\phi)}_{(k)}
\, , \label{phiEOMofLagPhikAl} \\
\Theta^\mu&=& \Theta_{(R)}^\mu+\Theta_{(\phi)}^\mu+\Theta_{(M)}^\mu
\, . \label{SurftofLagPhikAl}
\eea
In the above equations, for the expressions
$\mathcal{E}^{(gr)}_{(M)\mu\nu}$ and $\mathcal{E}^{(\phi)}_{(k)}$,
as well as the boundary term $\Theta_{(M)}^\mu$, see
Eq. (\ref{EOMofLagranLM}). Both the $\mathcal{E}^{(gr)}_{(\phi)\mu\nu}$
and $\tilde{\mathcal{E}}^{(\phi)}_{(k)}$ expressions of the field
equations and the surface term $\Theta_{(\phi)}^\mu$ are presented by
Eqs. (\ref{EOMforLphi}) and (\ref{SurftermLphi}) respectively.
In addition, the expression for the equation of motion
$\mathcal{E}^{(gr)}_{(R)\mu\nu}$ and the boundary term
$\Theta_{(R)}^\mu$, associated with the Lagrangian $\mathcal{L}_R$,
are read off as
\bea
\mathcal{E}^{(gr)}_{(R)\mu\nu}&=&G_{\mu\nu}
=R_{\mu\nu}-\frac{1}{2}g_{\mu\nu}R
\, , \nn \\
\Theta_{(R)}^\mu&=&2g^{\mu[\rho}g^{\nu]\sigma}\nabla_\nu\delta g_{\rho\sigma}
=\nabla^\mu(g_{\rho\sigma}\delta g^{\rho\sigma})
-\nabla_\rho \delta g^{\rho\mu}
\, . \label{EoMandSurftofGR}
\eea

According to Eq. (\ref{Genoffshcurr}), the off-shell Noether current with
respect to the Lagrangian (\ref{LagranPhikAl}) is defined by
\bea
J^\mu&=&2\mathcal{E}^{\mu\nu}\zeta_\nu+\zeta^\mu L
-\Theta^\mu\big(\mathcal{L}_\zeta g,\mathcal{L}_\zeta \phi^{(k)},
\mathcal{L}_\zeta A_{(p)}\big)
\nn \\
&=&J^\mu_{(R)}+J^\mu_{(\phi)}+J^\mu_{(M)}
\, . \label{OffshcurofLRMph}
\eea
In Eq. (\ref{OffshcurofLRMph}), the $J^\mu_{(R)}$ current, corresponding
to the Lagrangian $\mathcal{L}_R$, is defined through
\be
J^\mu_{(R)}=2\mathcal{E}^{(gr)\mu\nu}_{(R)}\zeta_\nu+\zeta^\mu L_R
-\Theta_{(R)}^\mu(\mathcal{L}_\zeta g)
\, . \label{OffshelcurrofGR}
\ee
Its corresponding potential $K_{(R)}^{\mu\nu}$ has the form
\be
K_{(R)}^{\mu\nu}=2\nabla^{[\mu}\zeta^{\nu]}
\, . \label{OffshelpotenofGR}
\ee
the $J^\mu_{(\phi)}$ current given by Eq. (\ref{OffshcurrLphi})
takes a zero value, that is, $J^\mu_{(\phi)}=0$, and the $J^\mu_{(M)}$ current,
associated with the Lagrangian $\mathcal{L}_M$, is presented by
(\ref{OffshcurrLM}). The quantity $\mathcal{E}^{\mu\nu}$ is given by
\bea
\mathcal{E}^{\mu\nu}&=&
\mathcal{E}^{(gr)\mu\nu}_{(R)}+\mathcal{E}^{(gr)\mu\nu}_{(\phi)}
+\mathcal{E}^{\mu\nu}_{(M)}
\, , \label{EmumuforLRPm}
\eea
where the specific expression for $\mathcal{E}^{\mu\nu}_{(M)}$ can be
found in Eq. (\ref{EMforLM}). The Noether current $J^\mu$ in
Eq. (\ref{OffshcurofLRMph}) further yields the off-shell Noether
potential
\bea
K_{R\phi M}^{\mu\nu}&=&K_{(R)}^{\mu\nu}+K_{(\phi)}^{\mu\nu}
+K_{(M)}^{\mu\nu}
\nn \\
&=&2\nabla^{[\mu}\zeta^{\nu]}
-\sum_{p=1}^{n-1}p(p+1)
U_{(p+1)}^{\mu\nu\mu_1\cdot\cdot\cdot\mu_{(p-1)}}
\zeta^\sigma A_{(p)\sigma\mu_1\cdot\cdot\cdot\mu_{(p-1)}}
\, , \label{OffsheNoepoLRPm}
\eea
where $K_{(\phi)}^{\mu\nu}=0$ and the potential $K_{(M)}^{\mu\nu}$,
associated with the Lagrangian $\mathcal{L}_M$, is determined by
Eq. (\ref{OffshpotenLM}).

Next, like before, supposing that the spacetime admits a Killing vector
$\xi^\mu$ that obeys all the conditions demanded in Eq. (\ref{GenKillvec}),
the off-shell ADT potential $Q_{R\phi M}^{\mu\nu}$ involving this
Killing vector is given in terms of the Noether potential
$K_{R\phi M}^{\mu\nu}$ in Eq. (\ref{OffsheNoepoLRPm}) and the
$\Theta^\mu\big(\delta g,\delta\phi^{(k)},\delta A_{(p)}\big)$
surface term in Eq. (\ref{SurftofLagPhikAl}) by
\bea
Q_{R\phi M}^{\mu\nu}
&=&\frac{1}{2}\frac{1}{\sqrt{-g}}
\delta\big(\sqrt{-g}K_{R\phi M}^{\mu\nu}(\xi)\big)
-\xi^{[\mu}\Theta^{\nu]}
\big(\delta g,\delta\phi^{(k)},\delta A_{(p)}\big)
\nn \\
&=&Q_{(R)}^{\mu\nu}+Q_{(\phi)}^{\mu\nu}
+Q_{(M)}^{\mu\nu}
\, , \label{ADTpotQofLRPm}
\eea
where the total potential $Q_{R\phi M}^{\mu\nu}$ is accordingly split
into three parts, i.e., the potentials
$Q_{(R)}^{\mu\nu}$, $Q_{(\phi)}^{\mu\nu}$ and
$Q_{(M)}^{\mu\nu}$, corresponding to the contributions from the
Lagrangian $\mathcal{L}_R$, $\mathcal{L}_\phi$ and  $\mathcal{L}_M$
respectively, are given by
\bea
Q_{(R)}^{\mu\nu}&=&\frac{1}{2}\frac{1}{\sqrt{-g}}
\delta\big(\sqrt{-g}K_{(R)}^{\mu\nu}(\xi)\big)
-\xi^{[\mu}\Theta_{(R)}^{\nu]}(\delta g)
\, , \label{OffsheADTpotR} \\
Q_{(\phi)}^{\mu\nu}&=&-\xi^{[\mu}\Theta_{(\phi)}^{\nu]}
\big(\delta\phi^{(k)}\big)
\, , \label{OffsheADTpotphi} \\
Q_{(M)}^{\mu\nu}&=&\frac{1}{2}\frac{1}{\sqrt{-g}}
\delta\big(\sqrt{-g}K_{(M)}^{\mu\nu}(\xi)\big)
-\xi^{[\mu}\Theta_{(M)}^{\nu]}\big(\delta A_{(p)}\big)
\, . \label{OffsheADTpotM}
\eea
The off-shell ADT current $\mathcal{J}_{R\phi M}^\mu$ for the
Lagrangian (\ref{LagranPhikAl}) can be got through the relation
$\mathcal{J}_{R\phi M}^\mu=\nabla_\nu Q_{R\phi M}^{\mu\nu}$.
When $\mathbb{L}_M$ in Eq. (\ref{LMform1anoth}) is adopted to take
the place of the quantity $L_M$ of the Lagrangian $\mathcal{L}_M$ in
the total Lagrangian (\ref{LagranPhikAl}), the ADT potential
$Q_{(M)}^{\mu\nu}$ can be substituted by $Q_{(\mathbb{L}_M)}^{\mu\nu}$,
which is defined by
\be
Q_{(\mathbb{L}_M)}^{\mu\nu}=\frac{1}{2}\frac{1}{\sqrt{-g}}
\delta\big(\sqrt{-g}K_{(\mathbb{L}_M)}^{\mu\nu}(\xi)\big)
-\xi^{[\mu}\Theta_{(\mathbb{L}_M)}^{\nu]}\big(\delta A_{(p)}\big)
\, , \label{OffsheADTforLM1anoth}
\ee
where the surface term $\Theta_{(\mathbb{L}_M)}^{\mu}$ and
the off-shell Noether potential $K_{(\mathbb{L}_M)}^{\mu\nu}$
can be found in Eqs. (\ref{SurftLagranLM1anoth}) and
(\ref{OffshelNopeteLM1anoth}) respectively. Furthermore, substituting
the off-shell ADT potential (\ref{ADTpotQofLRPm}) into
Eq. (\ref{QdefineAn}), one obtains the formula for the conserved charge
associated with the Lagrangian (\ref{LagranPhikAl}).

As we have mentioned in the previous section, the off-shell ADT potential
(\ref{ADTpotQofLRPm}) is essentially equivalent with those derived through
the LIW method \cite{LeeWald,IyerWald,IWaldentro} and
the BBC method \cite{BarnichB,Barnich,BarnichC,BCintegC} respectively.
For example, when Eq. (\ref{ADTpotQofLRPm}) is applied to compute the
off-shell ADT potential for the gravity theory with a single $p$-form
potential given in the works \cite{Rogatpform} and \cite{Compepform},
where potentials for this theory were derived through the LIW approach
and the BBC method respectively, one finds that all the potentials agree
with each other. On the other hand, as a consequence of the generality
for the Lagrangian $\mathcal{L}_M$, the potential (\ref{ADTpotQofLRPm})
can be widely used to compute the conserved charges of gravity theories
in the presence of scalar fields and $p$-form gauge fields. Its typical
applications are to calculate the conserved charges of black holes in
Einstein-Maxwell-dilaton theory and supergravity theories
\cite{ChowComp,Wu3CBH,TomMizog,GodelBHWuGi,ChongCLP,WuKKsdsBH}, such as
the dyonic AdS black holes in four-dimensional maximal $\mathcal{N}=8$,
$SO(8)$ gauged supergravity \cite{ChowComp}, the
general nonextremal rotating charged AdS black holes in five-dimensional
$U(1)^3$ gauged supergravity \cite{Wu3CBH}, the rotating charged
Kaluza-Klein black holes \cite{TomMizog} and the rotating charged
G\"{o}del-type black holes \cite{GodelBHWuGi} in five-dimensional minimal
supergravity, the general nonextremal charged rotating black holes in
five-dimensional minimal gauged supergravity \cite{ChongCLP}, and so on.
If the formula (\ref{QdefineAn}), endowed with the off-shell ADT potential
(\ref{ADTpotQofLRPm}), is applied to compute the conserved charges within
the low-energy effective field theory describing heterotic string theory
like in \cite{JJPIJMpA}, one can observe that all the results in that work
are covered.

\section{Conserved charges of (D+1)-dimensional Einstein gravity
under Kaluza-Klein reduction}\label{secthree}

In this section, we investigate the behaviour of conserved charges
in $(D+1)$-dimensional general relativity, which is defined through the
off-shell generalized ADT potential, by performing a Kaluza-Klein
reduction procedure along a compactified direction with $S^1$ topology.
As is known to us, the $(D+1)$-dimensional Einstein gravity is
described by the Einstein-Hilbert Lagrangian
\be
\hat{\mathcal{L}}_{\hat{R}}=\sqrt{-\hat{g}}\hat{R}
\, . \label{EinHilLagran}
\ee
Here and in what follows, all the quantities with the hat `` $^{\hat{}}$ ''
correspond to the ones in $(D+1)$-dimensional spacetime. It is assumed that the
$(D+1)$-dimensional spacetime manifold $M^{D+1}$ is the direct product of a
$D$-dimensional spacetime manifold
$M^{D}$ and an $S^1$ circle, namely, $M^{D+1}=M^{D}\otimes S^1$. The
$(D+1)$-dimensional spacetime may be endowed with the coordinate system
$\hat{x}^{\hat{\mu}}=(x^\mu, z)$, where $x^\mu$, parametrizing the manifold $M^{D}$,
runs over all the coordinates in $D$ dimensions and the $z$ coordinate
indicates the compactified direction on the $S^1$ circle of radius $L$. By further
introducing some $D$-dimensional fields, which are the metric tensor $g_{\mu\nu}$,
the Kaluza-Klein vector $\mathcal{A}_\mu$ and the dilaton $\varphi$ respectively,
the $(D+1)$-dimensional metric ansatz is supposed to take the following form
\cite{PopeKKth}:
\bea
d\hat{s}^2_{(D+1)}&=&e^{2\alpha\varphi}ds^2_{(D)}
+e^{2\beta\varphi}(dz+\mathcal{A})^2
\, , \nn \\
ds^2_{(D)}&=& g_{\mu\nu}dx^\mu dx^\nu
\, , \label{Dplus1metric}
\eea
where the $1$-form gauge field reads $\mathcal{A}=\mathcal{A}_\mu dx^\mu$, while
both the two dimensionally dependent constants $\alpha$ and $\beta$ are given by
\be
\beta=-(D-2)\alpha \, , \qquad
\alpha^2=\frac{1}{2(D-1)(D-2)}
\, . \label{alpbetconst}
\ee
It is supposed that all the components of the $(D+1)$-dimensional metric tensor
$\hat{g}_{\hat{\mu}\hat{\nu}}$ do not include the $z$ coordinate as a variable.
As a result, all the $D$-dimensional fields, such as $g_{\mu\nu}$, $\mathcal{A}_\mu$
and $\varphi$, are independent of the coordinate $z$. By performing a
Kaluza-Klein dimensional reduction of the $(D+1)$-dimensional
ansatz (\ref{Dplus1metric}) along the $z$ direction, the $(D+1)$-dimensional
Einstein-Hilbert Lagrangian (\ref{EinHilLagran}), which only consists of the
$(D+1)$-dimensional metric tensor $\hat{g}_{\hat{\mu}\hat{\nu}}$, can be reexpressed
in terms of all the $D$-dimensional fields as
\bea
\hat{\mathcal{L}}_{\hat{R}}&=&\mathcal{L}_{EMD}
=\sqrt{-g}L_{EMD}
\nn \\
&=&\sqrt{-g}\Big(
R-\frac{1}{2}\nabla^\mu\varphi\nabla_\mu\varphi
-\frac{1}{4}e^{-2(D-1)\alpha\varphi}\mathcal{F}^2\Big)
\, , \label{DdemLagran}
\eea
where the 2-form field strength $\mathcal{F}=d\mathcal{A}$. Note that we
have dropped the $\Box\varphi$ term that makes no contribution to the
field equations and one can strictly prove that this term also contributes
nothing to the off-shell ADT potential. The lower-dimensional
Lagrangian $\mathcal{L}_{EMD}$ is the one describing the
Einstein-Maxwell-dilaton theory for a particular value $-2(D-1)\alpha$
of dilaton coupling. Equation (\ref{DdemLagran}) demonstrates that the
$(D+1)$-dimensional Einstein gravity is essentially equivalent to the
$D$-dimensional theory equipped with a gravitational field, a scalar
and a $U(1)$ gauge field.

Letting $\hat{\xi}^{\hat{\mu}}$ denote a $(D+1)$-dimensional Killing
vector, according to Eq. (\ref{OffsheADTpotR}), an off-shell ADT potential
for the Einstein-Hilbert Lagrangian $\hat{\mathcal{L}}_{\hat{R}}$ takes the form
\bea
\hat{Q}_{(\hat{R})}^{\hat{\mu}\hat{\nu}}&=&
\frac{1}{2}\frac{1}{\sqrt{-\hat{g}}}
\delta\big(\sqrt{-\hat{g}}
\hat{K}_{(\hat{R})}^{\hat{\mu}\hat{\nu}}(\hat{\xi})\big)
-\hat{\xi}^{[\hat{\mu}}
\hat{\Theta}_{(\hat{R})}^{\hat{\nu}]}(\delta \hat{g})
\, . \label{Dplus1ADTpotGR}
\eea
On the other hand, with
help of Eq. (\ref{ADTpotQofLRPm}), the ADT potential related to the
$D$-dimensional Lagrangian $\mathcal{L}_{EMD}$ is given by
\bea
Q_{EMD}^{\mu\nu}
&=&\frac{1}{2}\frac{1}{\sqrt{-g}}
\delta\big(\sqrt{-g}K_{EMD}^{\mu\nu}(\xi)\big)
-\xi^{[\mu}\Theta_{EMD}^{\nu]}
\big(\delta g,\delta\varphi,\delta \mathcal{A}\big)
\, , \label{ADTpotQofLEMD}
\eea
where the $K_{EMD}^{\mu\nu}$ potential, which can be deduced from
Eq. (\ref{OffsheNoepoLRPm}), and the surface term
$\Theta_{EMD}^{\mu}$ are presented by
\bea
K_{EMD}^{\mu\nu}&=&
2\nabla^{[\mu}\xi^{\nu]}
+e^{-2(D-1)\alpha\varphi}\xi^\sigma\mathcal{A}_\sigma
\mathcal{F}^{\mu\nu}
\, , \nn \\
\Theta_{EMD}^{\mu}&=&
\nabla^\mu(g_{\rho\sigma}\delta g^{\rho\sigma})
-\nabla_\rho \delta g^{\rho\mu}
\nn \\
&&-(\nabla^\mu\varphi)\delta \varphi
-e^{-2(D-1)\alpha\varphi}\mathcal{F}^{\mu\nu}\delta\mathcal{A}_\nu
\, , \label{KThetaEMD}
\eea
respectively. The vector $\xi^\mu$ in Eq. (\ref{KThetaEMD}) is the
$D$-dimensional Killing vector, reflecting the symmetry of spacetime
and satisfying
\be
\mathcal{L}_\xi g_{\mu\nu} =0 \, , \quad
\mathcal{L}_\xi \varphi =0 \, , \quad
\mathcal{L}_\xi \mathcal{A} =0
\, . \label{KillvecinGR}
\ee

In the Appendix \textbf{\ref{appendC}}, we have proved that the
$\hat{\xi}_{(b)}^{\hat{\mu}}$ vector, defined through
$\hat{\xi}_{(b)}^{\hat{\mu}}=(\xi^\mu,b)$, is a
Killing vector in $(D+1)$ dimensions, where the parameter $b$ is
an arbitrary constant, if the $D$-dimensional $\xi^\mu$ vector satisfies
Eq. (\ref{KillvecinGR}). The $\hat{\xi}_{(b)}^{\hat{\mu}}$ vector
is associated with the off-shell Noether potential
$\hat{K}_{(\hat{R})}^{\hat{\mu}\hat{\nu}}\big(\hat{\xi}_{(b)}\big)
=\hat{K}_{(\hat{R})}^{\hat{\mu}\hat{\nu}}\big(\hat{\xi}\rightarrow\hat{\xi}_{(b)}\big)$,
whose $(\mu,\nu)$-component is reexpressed in terms of all the
lower-dimensional fields as
\be
\hat{K}_{(\hat{R})}^{\mu\nu}\big(\hat{\xi}_{(b)}\big)=
e^{-2\alpha\varphi}K_{EMD}^{\mu\nu}
+2\xi^{[\mu}\nabla^{\nu]}e^{-2\alpha\varphi}
+be^{-2D\alpha\varphi}\mathcal{F}^{\mu\nu}
\, . \label{HatKintermKEMD}
\ee
To derive Eq. (\ref{HatKintermKEMD}), we have made use of
Eq. (\ref{Dplus1Killvecmun}). On the other hand, by utilizing
Eq. (\ref{Dplus1SuftGR2}), one finds that the $\mu$-component of the
$(D+1)$-dimensional surface term $\hat{\Theta}_{(\hat{R})}^{\hat{\mu}}$
and the $D$-dimensional one $\Theta_{EMD}^{\mu}$ has the following
relationship
\bea
\hat{\Theta}^{\mu}_{(\hat{R})}&=&
e^{-2\alpha\varphi}\Theta_{EMD}^{\mu}
+\frac{e^{-2\alpha\varphi}}{\sqrt{-g}}
\delta\big(\sqrt{-g}e^{2\alpha\varphi}\nabla^\mu e^{-2\alpha\varphi}\big)
\, . \label{HatSurftinSfEMD}
\eea
Hence, when
$\hat{\xi}_{(b)}^{\hat{\mu}}=\hat{\xi}_{(0)}^{\hat{\mu}}=(\xi^\mu,0)$
, Eqs. (\ref{HatKintermKEMD}) and
(\ref{HatSurftinSfEMD}) yield
\be
\sqrt{-\hat{g}}\hat{Q}_{(\hat{R})}^{\mu\nu}
\big(\hat{\xi}_{(0)}\big)=
\sqrt{-g}Q_{EMD}^{\mu\nu}
+\sqrt{-g}e^{2\alpha\varphi}\delta\xi^{[\mu}\nabla^{\nu]} e^{-2\alpha\varphi}
\, . \label{QmuninDanDplus1D}
\ee
In Eq. (\ref{QmuninDanDplus1D}), the $(\mu,\nu)$-component of the $(D+1)$-dimensional
off-shell ADT potential multiplied by the factor $\sqrt{-\hat{g}}$ coincides with
the $D$-dimensional ADT potential multiplied by the factor $\sqrt{-g}$, if the
variation of the Killing vector vanishes, that is, $\delta\xi^\mu=0$. This is always
guaranteed when the Killing vectors are associated with the mass and angular momentum.
On the other hand, when $\hat{\xi}_{(b)}^{\hat{\mu}}=\hat{\xi}_{(1)}^{\hat{\mu}}
=(0,\cdot\cdot\cdot,0,1)$, one gets
\be
\sqrt{-\hat{g}}\hat{Q}_{(\hat{R})}^{\mu\nu}
\big(\hat{\xi}_{(1)}\big)=\frac{1}{2}
\delta\Big(\sqrt{-g}e^{-2(D-1)\alpha\varphi}\mathcal{F}^{\mu\nu}
\Big)
\, . \label{QmuninDanDplus1D2}
\ee

On basis of Eq. (\ref{QmuninDanDplus1D}), together with help of the
formula (\ref{QdefineAn}) for the conserved charge and
the relationship between the $(D+1)$-dimensional gravitational
constant $\hat{G}_{(D+1)}$ and the one $G_{(D)}$ in $D$ dimensions, namely,
\be
\hat{G}_{(D+1)}=2\pi L G_{(D)}
\, , \label{Graconst}
\ee
where $L$ is the radius of the $S^1$ circle, one observes that the conserved
charges in $(D+1)$ dimensions coincide with the ones in $D$ dimensions. That
is to say, the conserved charge defined in terms of the off-shell generalized
ADT potential is invariant under Kaluza-Klein dimensional reduction in the
framework of the theory for Einstein gravity. It has been mentioned that the
off-shell generalized ADT method is equivalent with the LIW and BBC methods,
so the conserved charges defined via these methods also exhibit the property
that they are invariant
under Kaluza-Klein reduction along a compactified direction.
On the other hand, according to the formula (\ref{QdefineAn}),
the conserved charge associated with the Killing vector $\hat{\xi}_{(1)}^{\hat{\mu}}$
is defined by
\be
\mathcal{Q}_z=\frac{1}{32\pi (D-2)! G_{(D)}} \int_{\partial\Sigma}
e^{-2(D-1)\alpha\varphi}\mathcal{F}^{\mu\nu}
\epsilon_{\mu\nu\mu_1\mu_2\cdot\cdot\cdot\mu_{(D-2)}}dx^{\mu_1}\wedge\cdot\cdot\cdot
\wedge dx^{\mu_{(D-2)}}
\, , \label{QzofDplus1GR}
\ee
where $\partial\Sigma$ is the boundary of a $(D-1)$-dimensional
hypersurface $\Sigma$. The charge $\mathcal{Q}_z$ is nothing
but the electric charge with respect to the Kaluza-Klein vector
$\mathcal{A}_\mu$ in the $D$-dimensional Einstein-Maxwell-dilaton
theory described by the Lagrangian (\ref{DdemLagran}). As a consequence,
we conclude that the angular momentum along the $z$ direction in the
context of the higher-dimensional Einstein gravity theory
is just the electric charge associated with the Kaluza-Klein vector
in the lower-dimensional theory when the conserved charge is defined
in terms the off-shell ADT formalism.

\section{Properties for conserved charges in gravity theories with n-form
field strength under Kaluza-Klein reduction}\label{secfour}

In this section, we extend the analysis in the previous section to investigate
the properties of conserved charges for gravity theories including a gravitational
filed, a single $n$-form field strength and $m$ scalar field $\phi^{(k)}$
$(k=1,\cdot\cdot\cdot,m)$ by
performing Kaluza-Klein dimensional reduction along a compactified direction
on a circle. In comparison with the case of Einstein gravity, we wonder whether
the conserved charge will be influenced by the Kaluza-Klein reduction when the
matter fields are included.

Without loss of generality, we take into consideration the $(D+1)$-dimensional
Lagrangian that has the form
\bea
\hat{\mathcal{L}}_{\hat{R}\hat{F}}&=&
\hat{\mathcal{L}}_{\hat{R}}
+\hat{\mathcal{L}}_{\hat{F}}+\hat{\mathcal{L}}_\phi
\, , \nn \\
\hat{\mathcal{L}}_{\hat{F}}&=&\sqrt{-\hat{g}}
\gamma\big(\phi^{(k)}\big) \hat{F}^2_{(n)}
\, , \nn \\
\hat{\mathcal{L}}_\phi&=&
\sqrt{-\hat{g}}\Big[\sum^m_{i,j=1}X_{ij}\big(\phi^{(k)}\big)
\hat{\nabla}^{\hat{\mu}}\phi^{(i)}\hat{\nabla}_{\hat{\mu}}\phi^{(j)}
+V\big(\phi^{(k)}\big)\Big]
\, , \label{Lagranwithnform}
\eea
where the Lagrangian $\hat{\mathcal{L}}_{\hat{R}}$ is given in
Eq. (\ref{EinHilLagran}), and the $n$-form field strength
$\hat{F}_{(n)}=d\hat{A}_{(n-1)}$. What is more,
in Eq. (\ref{Lagranwithnform}),
the $(n-1)$-form potential $\hat{A}_{(n-1)}$ in $(D+1)$ dimensions
is defined in terms of the $D$-dimensional gauge fields
$A_{(n-1)}$ and $A_{(n-2)}$ by
\bea
\hat{A}_{(n-1)\mu_1\cdot\cdot\cdot\mu_{(n-1)}}&=&
A_{(n-1)\mu_1\cdot\cdot\cdot\mu_{(n-1)}}
\, , \nn \\
\hat{A}_{(n-1)\mu_1\cdot\cdot\cdot\mu_{(n-2)}z}&=&
A_{(n-2)\mu_1\cdot\cdot\cdot\mu_{(n-2)}}
\, . \label{Dplus1Anmione}
\eea
From Eq. (\ref{Dplus1Anmione}), one easily gets all the
components of the $\hat{F}_{(n)}$ field strength that are
\bea
\hat{F}_{(n)\mu_1\cdot\cdot\cdot\mu_{n}}&=&
F_{(n)\mu_1\cdot\cdot\cdot\mu_{n}}
\nn \\
&=&n\nabla_{[\mu_1}A_{(n-1)\mu_2\cdot\cdot\cdot\mu_{n}]}
\, , \nn \\
\hat{F}_{(n)\mu_1\cdot\cdot\cdot\mu_{(n-1)}z}&=&
F_{(n-1)\mu_1\cdot\cdot\cdot\mu_{(n-1)}}
\nn \\
&=&(n-1)\nabla_{[\mu_1}A_{(n-2)\mu_2\cdot\cdot\cdot\mu_{(n-1)}]}
\, . \label{Dplus1Fnstrength}
\eea
In the present case, the $(D+1)$-dimensional metric ansatz is assumed to
take the same form as the one in Eq. (\ref{Dplus1metric}). Besides,
all the fields, including the metric tensor $\hat{g}_{\hat{\mu}\hat{\nu}}$,
the $A_{(n-1)}$ and $A_{(n-2)}$ potentials, and the scalar fields
$\phi^{(k)}$, are assumed to have no dependence on the $z$ coordinate.
Obviously, both the lower-dimensional $F_{(n)}$ and $F_{(n-1)}$
strengths, together with the functions $X_{ij}$, $V$ and $\gamma$,
are independent of the $z$ coordinate as well.

Now, following the general derivation for the Lagrangian with multiple scalar
fields and $p$-form potentials in Appendix \textbf{\ref{appendD}} to perform
Kaluza-Klein dimensional reduction along the $z$ direction on basis of the
$(D+1)$-dimensional ansatz (\ref{Dplus1metric}), we further send the
Lagrangian (\ref{Lagranwithnform}) to the form
\bea
\hat{\mathcal{L}}_{\hat{R}\hat{F}}&=&
\mathcal{L}_{RF}
\nn \\
&=&\mathcal{L}_{EMD}+\mathcal{L}_{F}+\tilde{\mathcal{L}}_{\phi}
\, , \label{DdLagranwithnf}
\eea
where the quantities $\mathcal{L}_{F}$ and $\tilde{\mathcal{L}}_{\phi}$
are given by
\bea
\mathcal{L}_{F}&=&\sqrt{-g}
\gamma \big(e^{-2(n-1)\alpha\varphi}\tilde{F}^2_{(n)}
+ne^{2(D-n)\alpha\varphi}F^2_{(n-1)}\big)
\, , \nn \\
\tilde{\mathcal{L}}_\phi&=&
\sqrt{-g}\Big(\sum^m_{i,j=1}X_{ij}
\nabla^{\mu}\phi^{(i)}\nabla_{\mu}\phi^{(j)}
+e^{2\alpha\varphi}V\Big)
\, , \label{LFandTildLph}
\eea
and see Eq. (\ref{DdemLagran}) for the expression of $\mathcal{L}_{EMD}$.
In Eq. (\ref{LFandTildLph}), the $n$-form field strength $\tilde{F}_{(n)}$
is defined by
\be
\tilde{F}_{(n)\mu_1\cdot\cdot\cdot\mu_{n}}=
F_{(n)\mu_1\cdot\cdot\cdot\mu_{n}}
-nF_{(n-1)[\mu_1\cdot\cdot\cdot\mu_{(n-1)}}
\mathcal{A}_{\mu_{n}]}
\, . \label{TildeFnform}
\ee
Note that here $d\tilde{F}_{(n)}\neq 0$.

Next, we pay attention to calculating the off-shell ADT potentials for
the Lagrangians $\mathcal{L}_{F}$ and $\hat{\mathcal{L}}_{\hat{F}}$.
As before, we introduce a Killing vector $\xi^\mu$ in the $D$-dimensional
spacetime, which also satisfies the following equations
\bea
\mathcal{L}_\xi A_{(n-2)} &=&0 \, ,\qquad
\mathcal{L}_\xi A_{(n-1)} =0 \, , \nn \\
\mathcal{L}_\xi \phi^{(k)}&=&0 \, , \qquad
(k=1, \cdot\cdot\cdot, m)
\, , \label{LiedeAnman1}
\eea
apart from the ones in (\ref{KillvecinGR}). According to
Eq. (\ref{OffsheADTpotM}), the $\xi^\mu$ Killing vector is in one-to-one
correspondence with the ADT potential for the $D$-dimensional
Lagrangian $\mathcal{L}_{F}$, which takes the form
\be
Q_{(F)}^{\mu\nu}=\frac{1}{2}\frac{1}{\sqrt{-g}}
\delta\big(\sqrt{-g}K_{(F)}^{\mu\nu}(\xi)\big)
-\xi^{[\mu}\Theta_{(F)}^{\nu]}\big(\delta A_{(n-2)},\delta A_{(n-1)}\big)
\, , \label{ADTpotforDdLF}
\ee
in which the off-shell Noether potential $K_{(F)}^{\mu\nu}(\xi)$, defined
through Eq. (\ref{OffshpotenLM}), are read off as
\bea
K_{(F)}^{\mu\nu}(\xi)&=&-2n(n-1)\gamma e^{-2(n-1)\alpha\varphi}
\xi^\sigma\big\{\tilde{F}_{(n)}^{\mu\nu\mu_1\cdot\cdot\cdot\mu_{(n-2)}}\times
\nn \\
&&\big[A^{(n-1)}_{\sigma\mu_1\cdot\cdot\cdot\mu_{(n-2)}}
-(n-2)\mathcal{A}_{\mu_{(n-2)}}
A^{(n-2)}_{\sigma\mu_1\cdot\cdot\cdot\mu_{(n-3)}}\big]
\nn \\
&&+(n-2)e^{2(D-1)\alpha\varphi}
F_{(n-1)}^{\mu\nu\mu_1\cdot\cdot\cdot\mu_{(n-3)}}
A^{(n-2)}_{\sigma\mu_1\cdot\cdot\cdot\mu_{(n-3)}}
\big\}
\, , \label{DdoffshADTKF}
\eea
and the surface term $\Theta_{(F)}^{\mu}$, defined by
Eq. (\ref{EOMofLagranLM}), is expressed as
\bea
\Theta_{(F)}^{\mu}&=&2n\gamma e^{-2(n-1)\alpha\varphi}
\big\{\tilde{F}_{(n)}^{\mu\mu_1\cdot\cdot\cdot\mu_{(n-1)}}\times
\nn \\
&&\big[\delta A^{(n-1)}_{\mu_1\cdot\cdot\cdot\mu_{(n-1)}}
-(n-1)\mathcal{A}_{\mu_{(n-1)}}
\delta A^{(n-2)}_{\mu_1\cdot\cdot\cdot\mu_{(n-2)}}\big]
\nn \\
&&+(n-1)e^{2(D-1)\alpha\varphi}
F_{(n-1)}^{\mu\mu_1\cdot\cdot\cdot\mu_{(n-2)}}
\delta A^{(n-2)}_{\mu_1\cdot\cdot\cdot\mu_{(n-2)}}
\big\}
\, . \label{DdSurfforLF}
\eea
In the parallel analysis, with help of Eq. (\ref{OffsheADTpotM}), the
off-shell ADT potential for the $(D+1)$-dimensional Lagrangian
$\hat{\mathcal{L}}_{\hat{F}}$ is given by
\be
\hat{Q}_{(\hat{F})}^{\hat{\mu}\hat{\nu}}=
\frac{1}{2}\frac{1}{\sqrt{-\hat{g}}}
\delta\big(\sqrt{-\hat{g}}
\hat{K}_{(\hat{F})}^{\hat{\mu}\hat{\nu}}(\hat{\xi})\big)
-\hat{\xi}^{[\hat{\mu}}\hat{\Theta}_{(\hat{F})}^{\hat{\nu}]}
\big(\delta \hat{A}_{(n-1)}\big)
\, , \label{ADTpforDplu1dhatLF}
\ee
where
\bea
\hat{K}_{(\hat{F})}^{\hat{\mu}\hat{\nu}}(\hat{\xi})&=&-2n(n-1)\gamma
\hat{F}_{(n)}^{\hat{\mu}\hat{\nu}\hat{\mu}_1\cdot\cdot\cdot\hat{\mu}_{(n-2)}}
\hat{\xi}^{\hat{\sigma}}
\hat{A}^{(n-1)}_{\hat{\sigma}\hat{\mu}_1\cdot\cdot\cdot\hat{\mu}_{(n-2)}}
\, , \nn \\
\hat{\Theta}_{(\hat{F})}^{\hat{\mu}}&=&2n\gamma
\hat{F}_{(n)}^{\hat{\mu}\hat{\mu}_1\cdot\cdot\cdot\hat{\mu}_{(n-1)}}
\delta\hat{A}^{(n-1)}_{\hat{\mu}_1\cdot\cdot\cdot\hat{\mu}_{(n-1)}}
\, , \label{Dplus1SurfKF}
\eea
taking relatively simple forms.

As what has been done in the previous section, here we still expect to
get the exact relationship between the ADT potentials for the Lagrangians
$\mathcal{L}_{F}$ and $\hat{\mathcal{L}}_{\hat{F}}$. In order to do this,
we first compare the surface terms yielded by the variation of both the
Lagrangians. The $\mu$-component of the
$(D+1)$-dimensional surface term $\hat{\Theta}_{(\hat{F})}^{\hat{\mu}}$
is presented in terms of all the lower-dimensional fields by
\bea
\hat{\Theta}_{(\hat{F})}^{\mu}&=&2n\gamma\hat{g}^{\mu\nu}
\hat{F}_{(n)\nu}^{~~~~\hat{\mu}_1\cdot\cdot\cdot\hat{\mu}_{(n-1)}}
\hat{A}^{(n-1)}_{\hat{\mu}_1\cdot\cdot\cdot\hat{\mu}_{(n-1)}}
\nn \\
&&+2n\gamma\hat{g}^{\mu z}
\hat{F}_{(n)z}^{~~~~\hat{\mu}_1\cdot\cdot\cdot\hat{\mu}_{(n-1)}}
\hat{A}^{(n-1)}_{\hat{\mu}_1\cdot\cdot\cdot\hat{\mu}_{(n-1)}}
\nn \\
&=&e^{-2\alpha\varphi}\Theta_{(F)}^{\mu}
\, , \label{RelatofSurfF}
\eea
in which Eq. (\ref{Dplus1contrPP1}) in the Appendix \textbf{\ref{appendD}} has
been used in order to get the last identity. On the other hand,
the $(D+1)$-dimensional off-shell Nother potential
$\hat{K}_{(\hat{F})}^{\mu\nu}\big(\hat{\xi}_{(b)}\big)$,
where the $(D+1)$-dimensional Killing vector $\hat{\xi}_{(b)}^{\hat{\mu}}$
has the value $(\xi^\mu, b)$ and both the Eqs. (\ref{LiedeHatA1})
and (\ref{LiedeHatA2}) show that the Lie derivative of the
$\hat{A}_{(n-1)}$ potential along this
vector vanishes, is connected with the $D$-dimensional potential
$K_{(F)}^{\mu\nu}(\xi)$ through Eq. (\ref{HatKFintermKF}). Hence,
by combining Eq. (\ref{RelatofSurfF}) with Eq. (\ref{HatKFintermKF}),
one finds that
\be
\sqrt{-\hat{g}}\hat{Q}_{(\hat{F})}^{\mu\nu}\big(\hat{\xi}_{(0)}\big)
=\sqrt{-g}Q_{(F)}^{\mu\nu}
\,  \label{RelaHatQFandQF}
\ee
when the Killing vector
$\hat{\xi}_{(b)}^{\hat{\mu}}=\hat{\xi}_{(0)}^{\hat{\mu}}=(\xi^\mu,0)$.

Making use of (\ref{OffsheADTpotphi}), one observes that the ADT potentials
$\hat{Q}_{(\phi)}^{\mu\nu}$ and $\tilde{Q}_{(\phi)}^{\mu\nu}$,
associated with $\hat{\mathcal{L}}_\phi$ and $\tilde{\mathcal{L}}_\phi$
respectively, are related to each other through
\bea
\sqrt{-\hat{g}}\hat{Q}_{(\phi)}^{\mu\nu}\big(\hat{\xi}_{(0)}\big)
&=&\sqrt{-g}\tilde{Q}_{(\phi)}^{\mu\nu}\big(\hat{\xi}_{(0)}\big)
\nn \\
&=&-\sqrt{-g}\xi^{[\mu}\Theta_{(\phi)}^{\nu]}
\big(\delta\phi^{(k)}\big)
\, , \label{ADTpotfortilLpLphi}
\eea
in which the surface term $\Theta_{(\phi)}^{\mu}$ is presented in
Eq. (\ref{SurftermLphi}).
On basis of Eqs. (\ref{QmuninDanDplus1D}), (\ref{RelaHatQFandQF})
and (\ref{ADTpotfortilLpLphi}),
one further obtains the relationship between the total ADT potential
$\hat{Q}_{(\hat{R}\hat{F})}^{\mu\nu}$ for the higher-dimensional
Lagrangian $\mathcal{L}_{\hat{R}\hat{F}}$ and the one
$Q_{(RF)}^{\mu\nu}$ for the lower-dimensional Lagrangian
$\mathcal{L}_{RF}$, that is,
\bea
\sqrt{-\hat{g}}\hat{Q}_{(\hat{R}\hat{F})}^{\mu\nu}
\big(\hat{\xi}_{(0)}\big)&=&
\sqrt{-\hat{g}}\big[\hat{Q}_{(\hat{R})}^{\mu\nu}
\big(\hat{\xi}_{(0)}\big)
+\hat{Q}_{(\hat{F})}^{\mu\nu}\big(\hat{\xi}_{(0)}\big)
+\hat{Q}_{(\phi)}^{\mu\nu}\big(\hat{\xi}_{(0)}\big)
\big]
\nn \\
&=&\sqrt{-g}\big(Q_{EMD}^{\mu\nu}
+Q_{(F)}^{\mu\nu}
+\tilde{Q}_{(\phi)}^{\mu\nu}\big)
\nn \\
&=&\sqrt{-g}Q_{(RF)}^{\mu\nu}
\, . \label{HatQRFandQRF}
\eea
To arrive at the second equality in Eq. (\ref{HatQRFandQRF}), the variation
of the Killing vector $\xi^\mu$ is required to vanish since
$\sqrt{-\hat{g}}\hat{Q}_{(\hat{R})}^{\mu\nu}=\sqrt{-g}Q_{EMD}^{\mu\nu}$
holds under the condition that $\delta \xi^\mu=0$. Generally speaking,
the variation of the Killing vector with respect to mass or angular
momentum always disappears. As the case of Einstein gravity in the
previous section, from Eq. (\ref{HatQRFandQRF}), we are able to see that the
conserved charges defined through the formula (\ref{QdefineAn})
are unchanged before and after performing Kaluza-Klein reduction to
the gravity theory described by the Lagrangian (\ref{Lagranwithnform}).

Besides, when the Killing vector
$\hat{\xi}_{(b)}^{\hat{\mu}}=\hat{\xi}_{(1)}^{\hat{\mu}}
=(0,\cdot\cdot\cdot,0,1)$, with help of Eq. (\ref{HatKFintermKF}), the
off-shell ADT potential associated with this vector is read off as
\bea
\sqrt{-\hat{g}}\hat{Q}_{(\hat{R}\hat{F})}^{\mu\nu}
\big(\hat{\xi}_{(1)}\big)&=&-n(n-1)
\delta\Big(\sqrt{-g}\gamma e^{-2(n-1)\alpha\varphi}
\tilde{F}_{(n)}^{\mu\mu_1\cdot\cdot\cdot\mu_{(n-2)}\nu}
A^{(n-2)}_{\mu_1\cdot\cdot\cdot\mu_{(n-2)}}\Big)
\nn \\
&&+\frac{1}{2}
\delta\Big(\sqrt{-g}e^{-2(D-1)\alpha\varphi}\mathcal{F}^{\mu\nu}\Big)
\, . \label{HatQRFxi1}
\eea
In light of the formula (\ref{QdefineAn}),
the conserved charge associated with the potential (\ref{HatQRFxi1})
is defined by
\be
\tilde{\mathcal{Q}}_z=\mathcal{Q}_z
-\frac{n(n-1)}{8\pi G_{(D)}} \int_{\partial \Sigma}
\gamma e^{-2(n-1)\alpha\varphi}
\tilde{F}_{(n)}^{\mu\mu_1\cdot\cdot\cdot\mu_{(n-2)}\nu}
A^{(n-2)}_{\mu_1\cdot\cdot\cdot\mu_{(n-2)}}
d\Sigma_{\mu\nu}
\, , \label{tildQzofDplus1LM}
\ee
where the charge $\mathcal{Q}_z$ is given by Eq. (\ref{QzofDplus1GR}). The charge
$\tilde{\mathcal{Q}}_z$ is the angular momentum along the $z$ coordinate
in $(D+1)$ dimensions. In analogy with the case of Einstein gravity in
the last section, from a $D$-dimensional perspective, $\tilde{\mathcal{Q}}_z$ can
be thought of as
the electric charge associated with the $U(1)$ gauge field $\mathcal{A}_\mu$
in the lower-dimensional reduced theory described by the Lagrangian
(\ref{DdLagranwithnf}).

It is worth noting that we have assumed that the
higher-dimensional geometry allows a compactified dimension to exist so that its metric
ansatz can be decomposed as the form in Eq. (\ref{Dplus1metric}),
in order to get the property that the off-shell generalized ADT charges remain the same
at mathematical level before and after Kaluza-Klein dimensional reduction
in the frame work of the Einstein gravity theory and the gravity theories with a
single $n$-form field strength. All the results demonstrate that the consistent
dimensional reduction can yields consistent conserved charges. Due to this, one is
able to simplify the calculations of the conserved charges via performing dimensional
reduction or lifting. Generally speaking, the higher-dimensional gravity theories
consist of less fields and the Lagrangian describing these theories is more compact.
Accordingly, the off-shell ADT potential for the Lagrangian is simpler. Consequently,
to simplify calculations, the conserved charges for the lower-dimensional theories can
be evaluated in higher dimensions by performing
dimensional lifting. This will be illustrated in the next section.

\section{An example: conserved charges of Kaluza-Klein black holes}
\label{secfive}

As an application of the general derivation in section \ref{secthree},
we shall explicitly evaluate the mass and angular momentum of
five-dimensional rotating Kaluza-Klein black holes found in
\cite{5DKKrochBH} in this section.
These black holes are the exact solution of pure Einstein gravity in
five dimensions, described by the Einstein-Hilbert Lagrangian
$\hat{\mathcal{L}}_{(5D)}=\sqrt{-\hat{g}}\hat{R}$. Under Kaluza-Klein
dimensional reduction from five dimensions to four dimensions, the
Kaluza-Klein black holes can be seen as the rotating black holes
with both electric and magnetic charges in the
four-dimensional Einstein-Maxwell-dilaton theory with a particular
dilaton coupling. According to the lower-dimensional Lagrangian
(\ref{DdemLagran}), the four-dimensional Lagrangian for this theory
takes the form
\be
\mathcal{L}_{(4D)}
=\sqrt{-g}\Big(
R-\frac{1}{2}\nabla^\mu\varphi\nabla_\mu\varphi
-\frac{1}{4}e^{-\sqrt{3}\varphi}\mathcal{F}^2\Big)
\, . \label{4DEMdLagran}
\ee

Within the context of the five-dimensional Einstein gravity, the rotating
Kaluza-Klein black hole solution is read off as \cite{5DKKrochBH}
\bea
d\hat{s}^2_{(5)}&=&\frac{H_2}{H_1}\big(dz+A_{(4d)}\big)^2-\frac{H_3}{H_2}
\Big(dt+\big[2J(r-m)+m^2a(pq)^{\frac{1}{2}}\big]\sin^2\theta \frac{d\phi}{H_3}\Big)^2
\nn \\
&&+H_1\Big(\frac{dr^2}{\Delta(r)}+d\theta^2
+\Delta(r)\sin^2\theta\frac{d\phi^2}{H_3} \Big)
\, ,\label{5DroKKBH}
\eea
where the compactified coordinate $z\sim z+2\pi L$, the functions
$\Delta(r)$, $H_1$, $H_2$ and $H_3$ are given by
\bea
\Delta(r)&=&r^2+m^2a^2-2mr \, , \quad
H_3=\Delta(r)-m^2a^2\sin^2\theta
\, , \nn \\
H_1&=&H_3+p(r-m)+\frac{2}{a}\sqrt{(p/q)}\big(J+a^2P_m Q_e\cos\theta\big)
\, , \nn \\
H_2&=&H_3+q(r-m)+\frac{2}{a}\sqrt{(q/p)}\big(J-a^2P_m Q_e\cos\theta\big)
\, , \label{H123Delta}
\eea
respectively, while the Kaluza-Klein vector $A_{(4d)}$ reads
\bea
A_{(4d)}&=&-\Big[Q_e(2r+p-2m)-a P_m\sqrt{(q^3/p)}\cos\theta\Big]\frac{dt}{H_2}
\nn \\
&&-\Big\{2P_m(H_2+m^2a^2\sin^2\theta)\cos\theta
\nn \\
&&-(pq)^{-\frac{1}{2}}Q_e\big[p^2r-mp(p-2m)+4qP_m^2\big]a\sin^2\theta\Big\}
\frac{d\phi}{H_2}
\, . \label{KKvecof5DBH}
\eea
In Eqs. (\ref{5DroKKBH}), (\ref{H123Delta}) and (\ref{KKvecof5DBH}), the
parameters $(a,m,p,q)$ are integral constants, while the parameters
$(J,P_m,Q_e)$ are presented by
\bea
J&=&\frac{\sqrt{pq}(pq+4m^2)}{4(p+q)}a
\, , \nn \\
P_m&=&\frac{1}{2}\sqrt{\frac{p(p^2-4m^2)}{p+q}} \, , \qquad
Q_e=\frac{1}{2}\sqrt{\frac{q(q^2-4m^2)}{p+q}}
\, , \label{JPQval}
\eea
where $P_m$ is the magnetic charge for the reduced four-dimensional black hole,
and it will be shown that the parameters $J$ and $Q_e$ denote the angular
momentum and electric charge respectively. It is worth noting that the parameter
$a$ in the metric given in \cite{5DKKrochBH} is substituted by $ma$ in the
present metric (\ref{5DroKKBH}), like in \cite{5DKKBHKeCFT}.

By performing dimensional reduction to the rotating black hole solution
(\ref{5DroKKBH}) from five dimensions to four dimensions, we get the four-dimensional
black hole solution
\bea
ds^2_{(4)}&=&-\frac{H_3}{\sqrt{H_1H_2}}
\Big(dt+[2J(r-m)+m^2a\sqrt{pq}]\sin^2\theta \frac{d\phi}{H_3}\Big)^2
\nn \\
&&+\sqrt{H_1H_2}\Big(\frac{dr^2}{\Delta(r)}+d\theta^2
+\frac{\Delta(r)}{H_3}\sin^2\theta d\phi^2\Big)
\, , \nn \\
e^{\varphi/\sqrt{3}}&=&\sqrt{H_1/H_2} \, , \qquad
\mathcal{A}=A_{(4d)}
\, , \label{4DKKbhsolu}
\eea
which is an exact solution corresponding to the Lagrangian (\ref{4DEMdLagran}).

In terms of the five-dimensional metric ansatz (\ref{5DroKKBH}), we now
compute the mass and angular momentum of the Kaluza-Klein black holes. The fluctuation
of the gravitational field is determined by the infinitesimal change of the parameters
$(m, a, p, q)$, that is,
\be
m\rightarrow m+dm \, , \quad
a\rightarrow a+da \, , \quad
p\rightarrow p+dp \, , \quad
q\rightarrow q+dq \, .
\ee
The Killing vectors associated with mass and the angular momentum along the $\phi$
direction are $\hat{\xi}^{\hat{\mu}}_{(t)}=(-1,0,0,0,0)$ and
$\hat{\xi}^{\hat{\mu}}_{(\phi)}=(0,0,0,1,0)$ respectively. With the help of Eq.
(\ref{Dplus1ADTpotGR}), the $(t,r)$ components of the off-shell ADT potentials
corresponding to these Killing vectors are given by
\bea
\sqrt{-\hat{g}}\hat{Q}^{tr}_{(KK)}\big(\hat{\xi}_{(t)}\big)&=&
d(p)+d(q)+\mathcal{O}\Big(\frac{1}{r}\Big)
\, , \nn \\
\sqrt{-\hat{g}}\hat{Q}^{tr}_{(KK)}\big(\hat{\xi}_{(\phi)}\big)&=&
4d(J)+\mathcal{O}\Big(\frac{1}{r}\Big)
\, . \label{QtrofmasanforKK}
\eea
By making use of the formula (\ref{QdefineAn}) for the conserved charges, the
mass $M$ and the angular memontum $J_\phi$ reads
\be
M=\frac{p+q}{4G_4} \, , \qquad
J_\phi=\frac{J}{G_4}
\, . \label{massanforKKBH}
\ee
Here the mass $M$ is identified with the ADM mass but different from the Komar mass,
which is $M_{Komar}=q/(4G_4)$.

On the other hand, it is completely feasible for us to evaluate the mass and
the angular momentum along the $\phi$ coordinate in terms of the four-dimensional
reduced metric (\ref{4DKKbhsolu}). They coincide with the ones obtained in
five dimensions respectively. This verifies the frontal conclusion that the
off-shell generalized ADT charges are invariant under Kaluza-Klein dimensional
reduction. However, in the four-dimensional case, apart from the gravitational
field, the gauge field and scalar field have to be taken into consideration,
so the calculations of the conserved charges are more involved. Moreover,
one can test that the angular momentum along the compactified
direction $z$ is equal to the electric charge $Q_e$ of the four-dimensional
black hole (\ref{4DKKbhsolu}).

\section{Conclusions and discussions}

In the present work, to provide another novel understanding of the conserved
charges for gravity theories, we investigate the conserved charges of generally
diffeomorphism invariant gravity theories with various matter fields,
particularly the ones with scalar fields and
$p$-form potentials, through the off-shell
generalized ADT formalism, as well as the properties of the conserved charges
under Kaluza-Klein dimensional reduction on a circle $S^1$.

First, we construct an off-shell generalized ADT current (\ref{OffShADTCurr})
within the framework of the generally diffeomorphism invariant gravity theories
described by the generic Lagrangian (\ref{GenLagran}) through the linear
combination of the two currents
in Eqs. (\ref{currecalJ1}) and (\ref{CurreLievarLag}). The former arises from
the variation of equation (\ref{Emunuxiidenti}), which is the total divergence
for the contraction of the vector $\zeta_\nu$ with the tensor $\mathcal{E}^{\mu\nu}$
made up of expressions of the field equations,
while the latter is deduced from the Lie derivative for the variation equation
(\ref{VarGenLagran}) of the Lagrangian with respect to the Killing vector. The
newly constructed ADT current, containing the term
$\mathcal{L}_\xi\Theta^\mu\big(\delta g,\delta \psi^{(r)}\big)$ and the terms
with the variation of the Killing vector, formally differs from those presented in
\cite{KimKY,CiteHJPY,JJPeng}. However, in essence, all the off-shell ADT currents
are equivalent since $\mathcal{L}_\xi\Theta^\mu\big(\delta g,\delta \psi^{(r)}\big)=0$
holds for generally diffeomorphism gravity theories when $\delta\xi^\mu=0$,
while the variation of the Killing
vectors associated with mass and angular momentum always disappears. The merits of
the ADT current (\ref{OffShADTCurr}) are that the procedure to derive this current
becomes simple and it makes a natural and practical construction for derivation of
its corresponding potential. Moving on to derive the off-shell ADT potential
(\ref{ADTpotenQ}), we further present the formula (\ref{QdefineAn}) for the conserved
charges associated with the Lagrangian (\ref{GenLagran}).

Second, we present a generic Lagrangian (\ref{LagranPhikAl}) that describes a
wide range of gravity theories consisting of a gravitational field, multiple
scalar fields and $p$-form potentials, thanks to the $L_M$ term of this
Lagrangian, which possesses the two general structures in Eqs. (\ref{LMform1}) and
(\ref{LMform2}) so that it incorporates a large range of terms made up of
scalar fields and $p$-form potentials for the Lagrangians in the context of various
gravity theories with these matter fields. Making use of the off-shell generalized
ADT formalism, we derive the off-shell Noether current (\ref{OffshcurofLRMph})
and ADT potential (\ref{ADTpotQofLRPm}) associated with the Lagrangian
(\ref{LagranPhikAl}). Furthermore, substituting the potential
(\ref{ADTpotQofLRPm}) into the formula (\ref{QdefineAn}), one can obtain the
formulation of conserved charges for this Lagrangian. As indicated before,
the Lagrangian (\ref{LagranPhikAl}) is very generic. Hence, the formulation of
conserved charges is applicable to supergravity theories, the
Einstein-Maxwell-dilaton theory, the low-energy effective field theory of
heterotic string theory and so on.

Third, we investigate the behaviour of the conserved charges for the theory
of Einstein gravity as well as the gravity theory with a single $p$-form potential
and multiple scalar fields in arbitrary dimensions by performing the Kaluza-Klein
dimensional reduction along a compactified direction. To both the two types of gravity
theories, as demonstrated by Eqs. (\ref{QmuninDanDplus1D}) and (\ref{HatQRFandQRF}),
the $(\mu,\nu)$ components of the $(D+1)$-dimensional off-shell ADT
potential multiplied by the factor $\sqrt{-\hat{g}}$ coincide with those
multiplied by the factor $\sqrt{-g}$ in $D$ dimensions. This directly leads to
the conclusion that the conserved charges defined in terms of the off-shell generalized
ADT formalism are invariant under Kaluza-Klein reduction on a circle. On the other hand,
the charges given by Eqs. (\ref{QzofDplus1GR}) and (\ref{tildQzofDplus1LM}) further
indicate that the angular momentum along the compactified direction in $(D+1)$ dimensions
is just the electric charge with respect to the Kaluza-Klein vector $\mathcal{A}_\mu$
in the $D$-dimensional reduced theory. What is more, it has been shown that the off-shell
generalized ADT method is essentially equivalent with the LIW and BBC methods. As a result,
the formulation of conserved charges defined on the basis of both the two methods naturally exhibit
the properties of the off-shell ADT formalism under Kaluza-Klein reduction. To illustrate
our calculations, we explicitly evaluate the mass and angular momentum of the five-dimensional
rotating Kaluza-Klein black hole (\ref{5DroKKBH}). These conserved charges coincide with
their corresponding ones for the reduced four-dimensional black hole (\ref{4DKKbhsolu}).

Besides, there are several issues deserving to be discussed and investigated in the
future work. We first take into consideration of the potential $\check{\mathcal{Q}}^{\mu\nu}$
given in Eq. (\ref{PotenchekQmm}), which arises from the linear combination of the
current $\mathcal{J}^\mu_{[1]}$ with the one $\mathcal{J}^\mu_{[2]}$ and depends on
a generic constant $k$. In the case where $k\neq 0$ and all the fields are off-shell,
further investigation is demanded to clarify whether this potential can be employed to
present well-defined and meaningful conserved charges of gravity theories, as well as
to understand the meaning of these conserved charges if it is possible. Next, the
formula (\ref{QdefineAn}) endowed with the potential (\ref{ADTpotQofLRPm}) can be
employed to derive the first law of thermodynamics for black holes with
$p$-form gauge fields and scalar fields.

Apart from the above two issues, as mentioned
before, we have analyzed the behaviour of the conserved charges for the Lagrangian
(\ref{Lagranwithnform}) that includes a sole $p$-form potential and multiple
scalar fields under Kaluza-Klein reduction. The results demonstrate that the conserved
charges for this Lagrangian are invariant before and after dimensional reduction
on a circle. As a matter of course, the analysis should be able to apply to the
more generic Lagrangian $\mathcal{L}_M$ with the function $L_M$ given by
Eq. (\ref{LMform1anoth}), which contains a number of $p$-form potentials.
Unfortunately, the calculations are very complex. However, we have checked several
other special cases of $\mathcal{L}_M$ and found that the conserved charges are
indeed invariant. As a consequence, we suppose that the conserved charges defined
in terms of the off-shell generalized ADT formalism are unchanged undergoing
Kaluza-Klein reduction along a compactified direction even for the Lagrangian
(\ref{LMform1anoth}) that consists of a lot of $p$-form potentials as well as
a number of scalar fields. Once this guess holds, one can further see that the
conserved charges associated with this Lagrangian are still unchanged under a
reduction from $(D+d)$ dimensions to $D$ dimensions on the $d$-torus
$T^d=S^1\times\cdot\cdot\cdot\times S^1$ because of the generality of the
Lagrangian. Another interesting issue is to prove that the conserved charges
via the off-shell generalized ADT formulation is invariant under Kaluza-Klein
reduction in the framework of generic covariant gravity theories involving
arbitrary matter fields and higher curvature terms.

\section*{Acknowledgments}

This work was supported by the Natural Science Foundation of China under Grant
No. 11505036 and No. 11275157. It was also partially supported by the Technology
Department of Guizhou province Fund under Grant No. (2016)1104 and the Guizhou
province science and technology innovation talent team [Grant No. (2015)4015].

\appendix
\section{The off-shell Noether currents and potentials of the Lagrangians
$\mathcal{L}_{M}$ and $\mathcal{L}_{\phi}$}\label{appendA}

In this appendix, we present detailed calculations for the derivations of
the off-shell Noether currents and potentials associated with the Lagrangians
$\mathcal{L}_M=\sqrt{-g} L_M$ and $\mathcal{L}_\phi=\sqrt{-g} L_\phi$ given
in Eq. (\ref{LagranPhikAl}). It has been shown in section \ref{sectwo} that
$L_M$ is assumed to take three types of general forms, i.e., the ones in
Eqs. (\ref{LMform1}), (\ref{LMform2}) and (\ref{LMform1anoth}). Accordingly,
our analysis associated with the Lagrangian $\mathcal{L}_M$ is divided into
three parts.

We first take into account the Noether current and potential for the Lagrangian
$\mathcal{L}_M$.
When $L_M$ takes the form as the one in Eq. (\ref{LMform1}), the variation
of the Lagrangian $\mathcal{L}_M$ is read off as
\bea
\delta \mathcal{L}_M &=&\sqrt{-g} \Big[\mathcal{E}^{(gr)}_{(M)\mu\nu}\delta g^{\mu\nu}
+\sum_{k=1}^m \mathcal{E}^{(\phi)}_{(k)}\delta\phi^{(k)}
+\sum_{p=1}^{n-1} \mathcal{E}_{(p)}^{(A)\mu_1\cdot\cdot\cdot\mu_p}
\delta A_{(p)\mu_1\cdot\cdot\cdot\mu_p}
\nn \\
&&+\nabla_\mu \Theta_{(M)}^\mu\big(\delta g,\delta\phi^{(k)},\delta A_{(p)}\big)\Big]
\, . \label{VarLagranLM}
\eea
Here the expressions of the field equations for the gravitational field, scalar fields
and the $p$-form potentials, as well as the surface term are given by
\bea
\mathcal{E}^{(gr)}_{(M)\mu\nu}&=&
\frac{\partial L_M}{\partial g^{\mu\nu}} -\frac{1}{2}g_{\mu\nu} L_M \, , \qquad
\mathcal{E}^{(\phi)}_{(k)}=\frac{\partial L_M}{\partial\phi^{(k)}}
\, , \nn \\
\mathcal{E}_{(p)}^{(A)\mu_1\cdot\cdot\cdot\mu_p}
&=&B_{(p)}^{\mu_1\cdot\cdot\cdot\mu_p}
-(p+1)\nabla_\mu U_{(p+1)}^{\mu\mu_1\cdot\cdot\cdot\mu_p}
\, , \nn \\
\Theta_{(M)}^\mu&=&\sum_{p=1}^{n-1}(p+1) U_{(p+1)}^{\mu\mu_1\cdot\cdot\cdot\mu_p}
\delta A_{(p)\mu_1\cdot\cdot\cdot\mu_p}
\, , \label{EOMofLagranLM}
\eea
respectively, where the two totally antisymmetric tensors
$B_{(p)}^{\mu_1\cdot\cdot\cdot\mu_p}$
and $U_{(p+1)}^{\mu_1\cdot\cdot\cdot\mu_{(p+1)}}$ are defined through
\bea
B_{(p)}^{\mu_1\cdot\cdot\cdot\mu_p}
&=&\frac{\partial L_M}{\partial A_{(p)\mu_1\cdot\cdot\cdot\mu_p}}
\, , \nn \\
U_{(p+1)}^{\mu_1\cdot\cdot\cdot\mu_{(p+1)}}
&=&\frac{\partial L_M}{\partial F_{(p+1)\mu_1\cdot\cdot\cdot\mu_{(p+1)}}}
\, , \label{DefinforBU}
\eea
respectively, and it is proved that the symmetric tensor
$\partial L_M/\partial g^{\mu\nu}$ satisfies
\be
\frac{\partial L_M}{\partial g^{\mu\nu}}=
\frac{1}{2}\sum_{p=1}^{n-1}\Big[p B_{(p)\mu}^{~~~~\mu_1\cdot\cdot\cdot\mu_{(p-1)}}
A_{(p)\nu\mu_1\cdot\cdot\cdot\mu_{(p-1)}}+
(p+1) U_{(p+1)\mu}^{~~~~~~~\mu_1\cdot\cdot\cdot\mu_p}
F_{(p+1)\nu\mu_1\cdot\cdot\cdot\mu_{p}}
\Big]
\, . \label{PartLMg}
\ee

The Lie derivative of $L_M$ along the $\zeta^\mu$ vector is
\bea
\zeta^\nu \nabla_\nu L_M&=&
\sum_{k=1}^m \mathcal{E}^{(\phi)}_{(k)}\zeta^\nu \nabla_\nu\phi^{(k)}
+\sum_{p=1}^{n-1} B_{(p)}^{\mu_1\cdot\cdot\cdot\mu_p}
\zeta^\nu \nabla_\nu A_{(p)\mu_1\cdot\cdot\cdot\mu_p}
\nn \\
&&+\sum_{p=1}^{n-1} U_{(p+1)}^{\mu_1\cdot\cdot\cdot\mu_{(p+1)}}
\zeta^\nu \nabla_\nu F_{(p+1)\mu_1\cdot\cdot\cdot\mu_{(p+1)}}
\, . \label{LieDeLM}
\eea
Substituting both the following equations
\bea
B_{(p)}^{\mu_1\cdot\cdot\cdot\mu_p}
\zeta^\nu \nabla_\nu A_{(p)\mu_1\cdot\cdot\cdot\mu_p}&=&
B_{(p)}^{\mu_1\cdot\cdot\cdot\mu_p}
\zeta^\nu F_{(p+1)\nu\mu_1\cdot\cdot\cdot\mu_{p}}
\nn \\
&&-p\Big(\nabla_\mu B_{(p)}^{\mu\mu_1\cdot\cdot\cdot\mu_{(p-1)}}\Big)
\zeta^\nu A_{(p)\nu\mu_1\cdot\cdot\cdot\mu_{(p-1)}}
\nn \\
&&+p\zeta^\nu \nabla_\mu\Big(B_{(p)}^{\mu\mu_1\cdot\cdot\cdot\mu_{(p-1)}}
A_{(p)\nu\mu_1\cdot\cdot\cdot\mu_{(p-1)}}\Big)
\nn \\
U_{(p+1)}^{\mu_1\cdot\cdot\cdot\mu_{(p+1)}}
\zeta^\nu \nabla_\nu F_{(p+1)\mu_1\cdot\cdot\cdot\mu_{(p+1)}}&=&
(p+1)U_{(p+1)}^{\mu\mu_1\cdot\cdot\cdot\mu_{p}}
\zeta^\nu \nabla_\mu F_{(p+1)\nu\mu_1\cdot\cdot\cdot\mu_{p}}
\eea
into Eq. (\ref{LieDeLM}), one arrives at the following
\bea
\zeta^\nu \nabla_\nu L_M&=&
\sum_{k=1}^m \mathcal{E}^{(\phi)}_{(k)}\zeta^\nu \nabla_\nu\phi^{(k)}
+\sum_{p=1}^{n-1} \mathcal{E}_{(p)}^{(A)\mu_1\cdot\cdot\cdot\mu_{p}}
\zeta^\nu F_{(p+1)\nu\mu_1\cdot\cdot\cdot\mu_{p}}
\nn \\
&&-\sum_{p=1}^{n-1}p\Big(\nabla_\mu
\mathcal{E}_{(p)}^{(A)\mu\mu_1\cdot\cdot\cdot\mu_{(p-1)}}\Big)
\zeta^\nu A_{(p)\nu\mu_1\cdot\cdot\cdot\mu_{(p-1)}}
+2\zeta ^\nu \nabla^\mu
\Big(\frac{\partial L_M}{\partial g^{\mu\nu}}\Big)
\, . \label{LieDeLM2}
\eea
On the other hand, for a $p$-form potential $A_{(p)}$, its Lie derivative
along the vector $\zeta^\mu$ can be defined as
\be
\mathcal{L}_\zeta A_{(p)\mu_1\cdot\cdot\cdot\mu_{p}} =
\zeta^\nu F_{(p+1)\nu\mu_1\cdot\cdot\cdot\mu_{p}}
-\sum_{k=1}^{p}(-1)^{k}\nabla_{\mu_k} \Big(\zeta^\nu
A_{(p)\nu\mu_1\cdot\cdot\cdot\mu_{(k-1)}\mu_{(k+1)}\cdot\cdot\cdot\mu_{p}}
\Big)
\, . \label{LiedeAl}
\ee
Making use of this definition, we have
\bea
\mathcal{E}_{(p)}^{(A)\mu_1\cdot\cdot\cdot\mu_p}
\mathcal{L}_\zeta A_{(p)\mu_1\cdot\cdot\cdot\mu_{p}}&=&
\mathcal{E}_{(p)}^{(A)\mu_1\cdot\cdot\cdot\mu_{p}}
\zeta^\nu F_{(p+1)\nu\mu_1\cdot\cdot\cdot\mu_{p}}
\nn \\
&&-p\Big(\nabla_\mu
\mathcal{E}_{(p)}^{(A)\mu\mu_1\cdot\cdot\cdot\mu_{(p-1)}}\Big)
\zeta^\nu A_{(p)\nu\mu_1\cdot\cdot\cdot\mu_{(p-1)}}
\nn \\
&&+p\nabla_\mu\Big(
\mathcal{E}_{(p)}^{(A)\mu\mu_1\cdot\cdot\cdot\mu_{(p-1)}}
\zeta^\nu A_{(p)\nu\mu_1\cdot\cdot\cdot\mu_{(p-1)}}\Big)
\, . \label{EAlLiedeAl}
\eea
With help of Eqs. (\ref{LieDeLM2}) and (\ref{EAlLiedeAl}), one can
obtain the identity
\bea
2\nabla_\mu \big(\mathcal{E}_{(M)}^{\mu\nu}\zeta_{\nu}\big)
&=&2\mathcal{E}_{(M)}^{(gr)\mu\nu}\nabla_\mu\zeta_{\nu}
-\sum_{k=1}^m \mathcal{E}^{(\phi)}_{(k)}\zeta^\nu \nabla_\nu\phi^{(k)}
\nn \\
&&-\sum_{p=1}^{n-1} \mathcal{E}_{(p)}^{(A)\mu_1\cdot\cdot\cdot\mu_p}
\mathcal{L}_\zeta A_{(p)\mu_1\cdot\cdot\cdot\mu_p}
\, ,\label{BianchforLM}
\eea
where the $\mathcal{E}_{(M)}^{\mu\nu}$ tensor is read off as
\be
\mathcal{E}_{(M)}^{\mu\nu}=
\mathcal{E}_{(M)}^{(gr)\mu\nu}-\frac{1}{2}
\sum_{p=1}^{n-1}p\mathcal{E}_{(p)}^{(A)\mu\mu_1\cdot\cdot\cdot\mu_{(p-1)}}
A^\nu_{(p)\mu_1\cdot\cdot\cdot\mu_{(p-1)}}
\, . \label{EMforLM}
\ee
It is shown in Eq. (\ref{EMforLM}) that the non-symmetric $2$-rank
tensor $\mathcal{E}_{(M)}^{\mu\nu}$ is only the combination for the
expressions of the field equations $\mathcal{E}_{(M)}^{(gr)\mu\nu}$
and $\mathcal{E}_{(p)}^{(A)\mu_1\cdot\cdot\cdot\mu_{p}}$,
which are associated with the gravitational field $g_{\mu\nu}$ and the gauge
fields $A_{(p)}$ respectively, while the contribution from the scalar fields
$\phi^{(k)}$ is absent. Equation (\ref{BianchforLM}) tests the identity
(\ref{generalizedBianch}) for generic covariant theories.
In particular, when the vector $\zeta^\mu$ becomes the
Killing vector $\xi^\mu$ that satisfies the conditions in Eq. (\ref{GenKillvec}),
Eq. (\ref{BianchforLM}) yields that the contraction between the tensor
$\mathcal{E}_{(M)}^{\mu\nu}$ and the Killing vector is divergence-free.

Following the procedure to derive the general off-shell Noether current
(\ref{Genoffshcurr}) in section \ref{secone}, one can further define an
off-shell Noether current $J_{(M)}^\mu$ with respect to the Lagrangian
$\mathcal{L}_{M}$ in terms of Eqs. (\ref{VarLagranLM}) and
(\ref{BianchforLM}) by
\bea
J_{(M)}^\mu&=&2\mathcal{E}_{(M)}^{\mu\nu}\zeta_\nu+\zeta^\mu L_M
-\Theta_{(M)}^\mu\big(\mathcal{L}_\zeta A_{(p)}\big)
\, , \nn \\
&=&\nabla_\nu K_{(M)}^{\mu\nu}
\, , \label{OffshcurrLM}
\eea
where the off-shell Noether potential $K_{(M)}^{\mu\nu}$ is presented
by
\be
K_{(M)}^{\mu\nu}=-\sum_{p=1}^{n-1}p(p+1)
U_{(p+1)}^{\mu\nu\mu_1\cdot\cdot\cdot\mu_{(p-1)}}
\zeta^\sigma A_{(p)\sigma\mu_1\cdot\cdot\cdot\mu_{(p-1)}}
\, , \label{OffshpotenLM}
\ee
which demonstrates that only the field strengths
$(F_{(q_1)},\cdot\cdot\cdot, F_{(q_J)})$ in $L_M$ are associated with the off-shell
Noether potential, while all the potentials $(A_{(p_1)},\cdot\cdot\cdot,A_{(p_I)})$
make no contribution to the potential.

On the other hand, in the case where $L_M$ takes the same form as the one
in Eq. (\ref{LMform2}), one only needs to postulate that
the Levi-Civita tensor density $\bar{\epsilon}_{\mu_1\cdot\cdot\cdot\mu_D}$ is
a $D$-form ``potential''
$A_{(D)\mu_1\cdot\cdot\cdot\mu_D}=\bar{\epsilon}_{\mu_1\cdot\cdot\cdot\mu_D}$.
Under such an assumption, one obtains that $F_{(D+1)}=dA_{(D)}=0$,
$\nabla_\mu\big(\sqrt{-g}A_{(D)\mu_1\cdot\cdot\cdot\mu_D}\big)=0$ and
$B_{(D)}^{\mu_1\cdot\cdot\cdot\mu_{D}}=0$ because of the disappearance of
variation for the $A_{(D)}$ potential. Due to these facts, one can further
observe that all the results with respect to $L_M$ described by Eq. (\ref{LMform1})
still hold in the case where $L_M$ possesses the form in
Eq. (\ref{LMform2}). This means that all the results are identified for both
the forms of $L_M$.

What is more, in the case where $L_M$ takes the form given in
Eq. (\ref{LMform1anoth}), that is $L_M=\mathbb{L}_M$,
all the results are equivalent with those for $L_M$ with the generic form in
Eq. (\ref{LMform1}). Therefore, here we do not plan to give detailed derivations
but directly present the surface term
$\Theta_{(\mathbb{L}_M)}^\mu$ and the off-shell Noether potential
$K_{(\mathbb{L}_M)}^{\mu\nu}$. The former, deduced from the surface term
$\Theta_{(M)}^\mu$ in Eq. (\ref{EOMofLagranLM}), has the form
\bea
\Theta_{(\mathbb{L}_M)}^\mu&=&\sum_{c=1}^{t}q_c
\mathbb{U}_{(q_c)}^{\mu\mu_1\cdot\cdot\cdot\mu_{(q_c-1)}}
\delta A_{(q_c-1)\mu_1\cdot\cdot\cdot\mu_{(q_c-1)}}
\nn \\
&&+\sum_{c=1}^{j}\tilde{q}_c
\mathbb{U}_{(\tilde{q}_c)}^{\mu\mu_1\cdot\cdot\cdot\mu_{(\tilde{q}_c-1)}}
\delta A_{(\tilde{q}_c-1)\mu_1\cdot\cdot\cdot\mu_{(\tilde{q}_c-1)}}
\, , \label{SurftLagranLM1anoth}
\eea
where the totally antisymmetric $q_c(\tilde{q}_c)$-rank tensors
$\mathbb{U}_{(q_c)}^{\mu_1\cdot\cdot\cdot\mu_{q_c}}$
and $\mathbb{U}_{(\tilde{q}_c)}^{\mu_1\cdot\cdot\cdot\mu_{\tilde{q}_c}}$
are defined by
the second equation in Eq. (\ref{DefinforBU}), namely,
$\mathbb{U}_{(q_c/\tilde{q}_c)}
=\partial\mathbb{L}_M/\partial F_{(q_c/\tilde{q}_c)}$.
However, it is worth noting that all the field strengths
$F_{(q_c/\tilde{q}_c)}$ are treated as independent variables when
the second equation in Eq. (\ref{DefinforBU}) is applied to get
$\mathbb{U}_{(q_c/\tilde{q}_c)}$. By following this, when $1\leq c\leq t$,
the tensor $\mathbb{U}_{(q_c)}$ reads
\bea
\mathbb{U}_{(q_c)}^{\mu_1\cdot\cdot\cdot\mu_{q_c}}&=&
W\mathbb{H}_{[c]\mu_{(q_c+1)}\cdot\cdot\cdot\mu_N}
Y_{(N)}^{\mu_1\cdot\cdot\cdot\mu_N}
\, , \nn \\
\mathbb{H}_{[c]\mu_{(q_c+1)}\cdot\cdot\cdot\mu_N}&=&
\big(F_{(q_1)}\cdot\cdot\cdot F_{(q_{(c-1)})}F_{(q_{(c+1)})}
\cdot\cdot\cdot F_{(q_t)}
A_{(p_1)}\cdot\cdot\cdot A_{(p_s)}\big)_{[\mu_{(q_c+1)}\cdot\cdot\cdot\mu_N]}
\, . \label{UforLM1anoth1}
\eea
In addition, when $1\leq c\leq j$, the tensor $\mathbb{U}_{(\tilde{q}_c)}$ is
given by
\bea
\mathbb{U}_{(\tilde{q}_c)}^{\mu_1\cdot\cdot\cdot\mu_{\tilde{q}_c}}&=&
W H_{(N)}^{\mu_1\cdot\cdot\cdot\mu_N}
\mathbb{Y}_{[c]\mu_{(\tilde{q}_c+1)}\cdot\cdot\cdot\mu_N}
\, , \nn \\
\mathbb{Y}_{[c]\mu_{(\tilde{q}_c+1)}\cdot\cdot\cdot\mu_N}&=&
\big(F_{(\tilde{q}_1)}\cdot\cdot\cdot F_{(\tilde{q}_{(c-1)})}F_{(\tilde{q}_{(c+1)})}
\cdot\cdot\cdot F_{(\tilde{q}_j)}
A_{(\tilde{p}_1)}\cdot\cdot\cdot A_{(\tilde{p}_i)}\big)_{[\mu_{(\tilde{q}_c+1)}\cdot\cdot\cdot\mu_N]}
\, . \label{UforLM1anoth2}
\eea
Starting out with Eq. (\ref{OffshpotenLM}), we derive the off-shell Noether potential
associated with $\mathbb{L}_M$, that is,
\bea
K_{(\mathbb{L}_M)}^{\mu\nu}&=&-\sum_{c=1}^{t}q_c(q_c-1)
\mathbb{U}_{(q_c)}^{\mu\nu\mu_1\cdot\cdot\cdot\mu_{(q_c-2)}}
\zeta^\sigma A_{(q_c-1)\sigma\mu_1\cdot\cdot\cdot\mu_{(q_c-2)}}
\nn \\
&&-\sum_{c=1}^{j}\tilde{q}_c(\tilde{q}_c-1)
\mathbb{U}_{(\tilde{q}_c)}^{\mu\nu\mu_1\cdot\cdot\cdot\mu_{(\tilde{q}_c-2)}}
\zeta^\sigma
A_{(\tilde{q}_c-1)\sigma\mu_1\cdot\cdot\cdot\mu_{(\tilde{q}_c-2)}}
\, . \label{OffshelNopeteLM1anoth}
\eea
Here the Noether potential $K_{(\mathbb{L}_M)}^{\mu\nu}$, as well as the
surface term in Eq. (\ref{SurftLagranLM1anoth}), is also applicable to
the case where $L_M$ has the same form as the one in Eq. (\ref{LMform2}).
For example, in such a case, if it is supposed that
$Y^{(D)}_{\mu_1\cdot\cdot\cdot\mu_D}=\bar{\epsilon}_{\mu_1\cdot\cdot\cdot\mu_D}$,
one finds that the surface term and the Noether potential are just the ones
given in Eqs. (\ref{SurftLagranLM1anoth}) and (\ref{OffshelNopeteLM1anoth})
in the absence of the $\mathbb{U}_{(\tilde{q}_c)}$ term, respectively.

In the following of this appendix, we calculate the off-shell Noether
current and potential of the Lagrangian $\mathcal{L}_{\phi}$. The variation
of the Lagrangian yields
\be
\delta \mathcal{L}_{\phi}=\sqrt{-g} \Big[\mathcal{E}^{(gr)}_{(\phi)\mu\nu}\delta g^{\mu\nu}
+\sum_{k=1}^m \tilde{\mathcal{E}}^{(\phi)}_{(k)}\delta\phi^{(k)}
+\nabla_\mu \Theta_{(\phi)}^\mu\big(\delta\phi^{(k)}\big)\Big]
\, , \label{VarLagranLphi}
\ee
where the expressions for the field equations $\mathcal{E}^{(gr)}_{(\phi)\mu\nu}$ and
$\tilde{\mathcal{E}}^{(\phi)}_{(k)}$ for the gravitational field and scalar fields
respectively, are given by
\bea
\mathcal{E}^{(gr)}_{(\phi)\mu\nu}&=&
\sum^m_{i,j=1}X_{ij}\nabla^\mu\phi^{(i)}\nabla_\mu\phi^{(j)}
-\frac{1}{2}g_{\mu\nu}L_\phi
\nn \\
\tilde{\mathcal{E}}^{(\phi)}_{(k)}&=&
-2\sum^m_{i=1}\nabla_\mu \big(X_{ik}\nabla^\mu\phi^{(i)}\big)
+\sum^m_{i,j=1}\frac{\partial X_{ij}}{\partial\phi^{(k)}}
\nabla^\mu\phi^{(i)}\nabla_\mu\phi^{(j)}
+\frac{\partial V}{\partial\phi^{(k)}}
\, , \label{EOMforLphi}
\eea
and the surface term $\Theta_{(\phi)}^\mu$ is read off as
\be
\Theta_{(\phi)}^\mu=
2\sum^m_{i,j=1}X_{ij}
\big(\nabla^\mu\phi^{(i)}\big)\delta\phi^{(j)}
\, . \label{SurftermLphi}
\ee
For the expressions for the equations of motion $\mathcal{E}^{(gr)}_{(\phi)\mu\nu}$
and $\tilde{\mathcal{E}}^{(\phi)}_{(k)}$, one is able to prove that they
satisfy the following identify
\be
2\nabla_\mu \big(\mathcal{E}_{(\phi)}^{\mu\nu}\zeta_{\nu}\big)
=2\mathcal{E}_{(\phi)}^{(gr)\mu\nu}\nabla_\mu\zeta_{\nu}
-\sum_{k=1}^m \tilde{\mathcal{E}}^{(\phi)}_{(k)}\zeta^\nu \nabla_\nu\phi^{(k)}
\, ,\label{BianchforLphi}
\ee
where $\mathcal{E}_{(\phi)}^{\mu\nu}=\mathcal{E}_{(\phi)}^{(gr)\mu\nu}$,
which contains no expressions of the field equations $\tilde{\mathcal{E}}^{(\phi)}_{(k)}$.
This implies that the scalar fields $\phi^{(k)}$ makes no contribution to the
divergence term, unlike the case for the gauge fields. Equivalently,
Eq. (\ref{BianchforLphi}) can be rewritten as
\be
\nabla_\mu \mathcal{E}_{(\phi)}^{(gr)\mu\nu}
+\frac{1}{2}\sum_{k=1}^m \tilde{\mathcal{E}}^{(\phi)}_{(k)} \nabla^\nu\phi^{(k)}=0
\, ,\label{BianchforLphi2}
\ee
which can be regarded as a generalized Bianchi identity for the scalar fields.
The same identity appears in the case of Horndeski theory \cite{JJPengPLB}.

As before, in terms of
the Eqs. (\ref{VarLagranLphi}) and (\ref{BianchforLphi}), we get the off-shell
Noether current for the Lagrangian $\mathcal{L}_{\phi}$ that has the form
\bea
J_{(\phi)}^\mu&=&2\mathcal{E}_{(\phi)}^{\mu\nu}\zeta_\nu+\zeta^\mu L_\phi
-\Theta_{(\phi)}^\mu\big(\mathcal{L}_\zeta \phi^{(k)}\big)
\nn \\
&=&0
\, , \label{OffshcurrLphi}
\eea
which yields a vanishing off-shell Noether potential $K_{(\phi)}^{\mu\nu}$,
namely,
\be
K_{(\phi)}^{\mu\nu}=0
\, . \label{OffshpotenLphi}
\ee
Nevertheless, in the case of the Horndeski theory, it was shown in \cite{JJPengPLB}
that the off-shell Noether potential related to the Lagrangian containing
the terms with higher derivatives of the scalar fields is non-zero.

\section{The (D+1)-dimensional Christoffel symbols and surface term}\label{appendB}

In this appendix, we first derive the explicit expressions of the
$(D+1)$-dimensional Christoffel symbols
$\hat{\Gamma}^{\hat{\rho}}_{\hat{\mu}\hat{\nu}}$ in terms of
all the $D$-dimensional fields, such as the gravitational field
$g_{\mu\nu}$, the Kaluza-Klein vector $\mathcal{A}_\mu$
and the dilaton $\varphi$. Subsequently the Christoffel symbols
are used to derive the exact relationship of the higher-dimensional
surface term
$\hat{\Theta}^{\hat{\mu}}_{(\hat{R})}$ to its lower-dimensional
counterpart.

We now calculate the Christoffel symbols
$\hat{\Gamma}^{\hat{\rho}}_{\hat{\mu}\hat{\nu}}$. On basis of the
$(D+1)$-dimensional metric ansatz (\ref{Dplus1metric}), the components of the
$(D+1)$-dimensional metric tensor $\hat{g}_{\hat{\mu}\hat{\nu}}$ are given
in terms of the $D$-dimensional fields by
\be
\hat{g}_{\mu\nu}=e^{2\alpha\varphi}g_{\mu\nu}
+e^{2\beta\varphi}\mathcal{A}_\mu\mathcal{A}_\nu \, , \quad
\hat{g}_{\mu z}=e^{2\beta\varphi}\mathcal{A}_\mu \, , \quad
\hat{g}_{zz}=e^{2\beta\varphi}
\, . \label{Dplus1gmunu}
\ee
As a result, the inverse $\hat{g}^{\hat{\mu}\hat{\nu}}$ of the metric tensor
are presented by
\be
\hat{g}^{\mu\nu}=e^{-2\alpha\varphi}g^{\mu\nu}
 \, , \quad
\hat{g}^{\mu z}=-e^{-2\alpha\varphi}\mathcal{A}^\mu \, , \quad
\hat{g}^{zz}=e^{-2\beta\varphi}
+e^{-2\alpha\varphi}\mathcal{A}_\mu\mathcal{A}^\mu
\, . \label{Dplus1gupmunu}
\ee
According to the definition for the Christoffel symbols, together
with help of the expressions of $\hat{g}_{\hat{\mu}\hat{\nu}}$ and
$\hat{g}^{\hat{\mu}\hat{\nu}}$, the components
$\hat{\Gamma}^{\rho}_{\mu\nu}$ and $\hat{\Gamma}^{z}_{\mu\nu}$ of
the higher-dimensional Christoffel symbols
$\hat{\Gamma}^{\hat{\rho}}_{\hat{\mu}\hat{\nu}}$ are given
in terms of the lower-dimensional fields by
\bea
\hat{\Gamma}^{\rho}_{\mu\nu}&=&
\Gamma^{\rho}_{\mu\nu}
+e^{-2\alpha\varphi}\delta^\rho_{(\mu}\nabla_{\nu)} e^{2\alpha\varphi}
-\frac{1}{2}e^{-2\alpha\varphi}g_{\mu\nu}\nabla^{\rho} e^{2\alpha\varphi}
\nn \\
&&+e^{2(\beta-\alpha)\varphi}\mathcal{A}_{(\mu}\mathcal{F}_{\nu)}^{~~\rho}
-\frac{1}{2}e^{-2\alpha\varphi}\mathcal{A}_{\mu}\mathcal{A}_{\nu}
\nabla^{\rho} e^{2\beta\varphi}
\, , \nn \\
\hat{\Gamma}^{z}_{\mu\nu}&=&
-\mathcal{A}_\rho\hat{\Gamma}^{\rho}_{\mu\nu}
+e^{-2\beta\varphi}\mathcal{A}_{(\mu}\nabla_{\nu)} e^{2\beta\varphi}
+\partial_{(\mu}\mathcal{A}_{\nu)}
\, , \label{Dplus1Gamrzmn}
\eea
where the field strength $\mathcal{F}_{\mu\nu}$ is defined through
$\mathcal{F}_{\mu\nu}=2\partial_{[\mu}\mathcal{A}_{\nu]}$, while the
other components are read off as
\bea
\hat{\Gamma}^{\rho}_{zz}&=&
-\frac{1}{2}e^{-2\alpha\varphi}\nabla^{\rho} e^{2\beta\varphi}
\, , \nn \\
\hat{\Gamma}^{z}_{zz}&=&
-\mathcal{A}_\rho\hat{\Gamma}^{\rho}_{zz}
=\frac{1}{2} e^{-2\alpha\varphi}\mathcal{A}_\rho\nabla^{\rho} e^{2\beta\varphi}
\, , \nn \\
\hat{\Gamma}^{\rho}_{\mu z}&=&
-\frac{1}{2}e^{-2\alpha\varphi}\mathcal{A}_\mu\nabla^{\rho} e^{2\beta\varphi}
+\frac{1}{2}e^{2(\beta-\alpha)\varphi}\mathcal{F}_{\mu}^{~\rho}
\, , \nn \\
\hat{\Gamma}^{z}_{\mu z}&=&
-\frac{1}{2}e^{2(\beta-\alpha)\varphi}
\mathcal{F}_{\mu}^{~\rho}\mathcal{A}_\rho
+\frac{1}{2}e^{-2\alpha\varphi}\mathcal{A}_\mu\mathcal{A}_\rho
\nabla^{\rho} e^{2\beta\varphi}
+\frac{1}{2}e^{-2\beta\varphi}\nabla_{\mu} e^{2\beta\varphi}
\, . \label{Dplus1Gammother}
\eea

Finally, utilizing the above higher-dimensional Christoffel symbols
$\hat{\Gamma}^{\hat{\rho}}_{\hat{\mu}\hat{\nu}}$, we evaluate the
$(D+1)$-dimensional surface term $\hat{\Theta}^{\hat{\mu}}_{(\hat{R})}$
in terms of all the $D$-dimensional fields. Its $\hat{\mu}=\mu$
components are presented by
\be
\hat{\Theta}^{\mu}_{(\hat{R})}=
\hat{\Theta}^{\mu}_{(\delta g)}
+\hat{\Theta}^{\mu}_{(\delta \mathcal{A})}
+\hat{\Theta}^{\mu}_{(\delta \phi)}
\, , \label{Dplus1SuftGR}
\ee
where
\bea
\hat{\Theta}^{\mu}_{(\delta g)}&=&
-e^{-2\alpha\varphi}[\nabla_\rho \delta g^{\rho\mu}
-\nabla^\mu(g_{\rho\sigma}\delta g^{\rho\sigma})]
-(D\alpha+\beta)e^{-2\alpha\varphi}(\nabla_\nu\varphi)\delta g^{\mu\nu}
\nn \\
&&+\alpha e^{-2\alpha\varphi}(\nabla^\mu\varphi)
g_{\rho\sigma}\delta g^{\rho\sigma}
\, , \nn \\
\hat{\Theta}^{\mu}_{(\delta \mathcal{A})}&=&
-e^{2(\beta-2\alpha)\varphi}\mathcal{F}^{\mu\nu}\delta\mathcal{A}_\nu
\, ,
\eea
and the $\hat{\Theta}^{\mu}_{(\delta \phi)}$ term is read off as
\bea
\hat{\Theta}^{\mu}_{(\delta \phi)}&=&
[2(D-1)\alpha-\beta](\nabla^\mu\varphi)\delta e^{-2\alpha\varphi}
+3\beta e^{2(\beta-\alpha)\varphi} (\nabla^\mu\varphi)\delta e^{-2\beta\varphi}
\nn \\
&&+(D-1)\nabla^\mu\big(\delta e^{-2\alpha\varphi}\big)
+e^{2(\beta-\alpha)\varphi}\nabla^\mu\big(\delta e^{-2\beta\varphi}\big)
\, .
\eea
Making use of the values of the $\alpha$ and $\beta$ constants in
Eq. (\ref{alpbetconst}), we simplify Eq. (\ref{Dplus1SuftGR}) as
\bea
\hat{\Theta}^{\mu}_{(\hat{R})}&=&
e^{-2\alpha\varphi}\big[\nabla^\mu(g_{\rho\sigma}\delta g^{\rho\sigma})
-\nabla_\rho \delta g^{\rho\mu}
-(\nabla^\mu\varphi)\delta \varphi
\nn \\
&&-e^{-2(D-1)\alpha\varphi}\mathcal{F}^{\mu\nu}\delta\mathcal{A}_\nu
\big]+ e^{-2\alpha\varphi}\big(\nabla^\mu e^{-2\alpha\varphi}\big)
\delta e^{2\alpha\varphi}
\nn \\
&&+\frac{1}{\sqrt{-g}}\delta\big(\sqrt{-g}\nabla^\mu e^{-2\alpha\varphi}\big)
\, . \label{Dplus1SuftGR2}
\eea
Here we do not plan to present the $\hat{\Theta}^{z}_{(\hat{R})}$ component
since the definition of the conserved charges does not involve this
quantity.

\section{The (D+1)-dimensional Killing vector}\label{appendC}

In the present appendix, the main goal consists in deriving
the Killing vector in $(D+1)$ dimensions on basis of the features
of the $D$-dimensional Killing vector, as well as studying the
behaviour of the Lie derivative for the gauge fields with respect to
the $(D+1)$-dimensional Killing vector.

We first prove that the $(D+1)$-dimensional
$\hat{\xi}_{(b)}^{\hat{\mu}}$ vector, defined through
\be
\hat{\xi}_{(b)}^{\hat{\mu}}=
(\hat{\xi}_{(b)}^{\mu}, \hat{\xi}_{(b)}^z)=(\xi^\mu, b)
\, , \label{Definhatximb}
\ee
is a Killing vector, where $\xi^\mu$ is the $D$-dimensional Killing
vector obeying Eq. (\ref{KillvecinGR}) and the parameter $b$ is an
arbitrary constant.

In order to prove $\hat{\xi}_{(b)}^{\hat{\mu}}$ is a Killing vector in
$(D+1)$-dimensional spacetime, we only need to prove the following
Killing equation
\be
\hat{\nabla}^{\hat{\mu}} \hat{\xi}_{(b)}^{\hat{\nu}}
+\hat{\nabla}^{\hat{\nu}} \hat{\xi}_{(b)}^{\hat{\mu}} =0
\, \label{Dplus1KillvecGR}
\ee
in local coordinates holds for any value of the vector
$\hat{\xi}_{(b)}^{\hat{\mu}}$. With help of
the Christoffel symbols in Appendix \textbf{\ref{appendB}}, we obtain
\bea
\hat{\nabla}^{\mu} \hat{\xi}_{(b)}^{\nu}&=&
e^{-2\alpha\varphi}\nabla^{\mu} \xi^{\nu}
+\frac{1}{2}e^{-4\alpha\varphi}g^{\mu\nu}
\xi^\sigma\nabla_\sigma e^{2\alpha\varphi}
+\xi^{[\mu}\nabla^{\nu]} e^{-2\alpha\varphi}
\nn \\
&&+\frac{1}{2}e^{2(\beta-2\alpha)\varphi}
(b+\mathcal{A}_\sigma \xi^\sigma)\mathcal{F}^{\mu\nu}
\, , \label{Dplus1Killvecmun}
\eea
which yields
\bea
\hat{\nabla}^{\mu} \hat{\xi}_{(b)}^{\nu}
+\hat{\nabla}^{\nu} \hat{\xi}_{(b)}^{\mu}&=&
-e^{-2\alpha\varphi} \mathcal{L}_\xi g^{\mu\nu}
-g^{\mu\nu}\mathcal{L}_\xi e^{-2\alpha\varphi}
\nn \\
&=& 0=- \mathcal{L}_\xi \hat{g}^{\mu\nu}
\, . \label{Dplus1Kivecmm}
\eea
The above equation implies that the $(\mu,\nu)$ components of
Eq. (\ref{Dplus1KillvecGR}) holds.

When $\hat{\mu}=\mu$ and $\hat{\nu}=z$, the left hand side of
Eq. (\ref{Dplus1KillvecGR}) is given in terms of the lower-dimensional
fields by
\bea
\hat{\nabla}^{\mu} \hat{\xi}_{(b)}^{z}
+\hat{\nabla}^{z} \hat{\xi}_{(b)}^{\mu}&=&
e^{-2\alpha\varphi}g^{\mu\rho}
\big(\mathcal{L}_\xi\mathcal{A}_\rho
-\mathcal{A}^\sigma\mathcal{L}_\xi g_{\rho\sigma}\big)
+\mathcal{A}^\mu \mathcal{L}_\xi e^{-2\alpha\varphi}
\nn \\
&=&0 =\mathcal{L}_\xi \hat{g}^{\mu z}
\, .\label{Dplus1Kivecmz}
\eea
Obviously, the $(\mu,z)$ components of Eq. (\ref{Dplus1KillvecGR})
also holds.

When $\hat{\mu}=z$ and $\hat{\nu}=z$, the left hand side of
Eq. (\ref{Dplus1KillvecGR}) is presented by
\bea
2\hat{\nabla}^{z} \hat{\xi}_{(b)}^{z}&=&
-\mathcal{L}_\xi \big(e^{-2\alpha\varphi}\mathcal{A}_\mu\mathcal{A}^\mu
+e^{-2\beta\varphi}\big)
\nn \\
&=& 0 = -\mathcal{L}_\xi \hat{g}^{zz}
\, , \label{Dplus1Kiveczz}
\eea
which means that the $(z,z)$-component obeys Eq. (\ref{Dplus1KillvecGR}). By a
combination of Eqs. (\ref{Dplus1Kivecmm}), (\ref{Dplus1Kivecmz}) and
(\ref{Dplus1Kiveczz}), one sees that the $(D+1)$-dimensional Killing equation
(\ref{Dplus1KillvecGR}) indeed holds for all the components.

Finally, we show that the Lie derivative of the potential
$\mathcal{L}_{\hat{\xi}} \hat{A}_{(n-1)\hat{\mu}_1\cdot\cdot\cdot\hat{\mu}_{(n-1)}}$
in $(D+1)$-dimensional spacetime along the Killing vector $\hat{\xi}^{\hat{\mu}}$
vanishes, that is,
\be
\mathcal{L}_{\hat{\xi}} \hat{A}_{(n-1)\hat{\mu}_1\cdot\cdot\cdot\hat{\mu}_{(n-1)}}
 =0
\, , \label{Dplus1DeriofAn}
\ee
if the Lie derivative of the potentials in $D$-dimensional spacetime
satisfies Eq. (\ref{LiedeAnman1}). In fact, according to the definition of the
Lie derivative, one sees that
\bea
\mathcal{L}_{\hat{\xi}} \hat{A}_{(n-1)\mu_1\cdot\cdot\cdot\mu_{(n-1)}}
&=&\hat{\xi}^{\hat{\mu}}\hat{\partial}_{\hat{\mu}}
\hat{A}_{(n-1)\mu_1\cdot\cdot\cdot\mu_{(n-1)}}
\nn \\
&&-\sum_{k=1}^{n-1}(-1)^{k}
\hat{A}_{(n-1)\hat{\nu}\mu_1\cdot\cdot\cdot\mu_{(k-1)}\mu_{(k+1)}\cdot\cdot\cdot\mu_{(n-1)}}
\hat{\partial}_{\hat{\mu}_k}\hat{\xi}^{\hat{\nu}}
\nn \\
&=&\xi^{\mu}\partial_\mu
A_{(n-1)\mu_1\cdot\cdot\cdot\mu_{(n-1)}}
\nn \\
&&-\sum_{k=1}^{n-1}(-1)^{k}
A_{(n-1)\nu\mu_1\cdot\cdot\cdot\mu_{(k-1)}\mu_{(k+1)}\cdot\cdot\cdot\mu_{(n-1)}}
\partial_{\mu_k}\xi^{\nu}
\nn \\
&=&\mathcal{L}_{\xi} A_{(n-1)\mu_1\cdot\cdot\cdot\mu_{(n-1)}}
=0 \, . \label{LiedeHatA1}
\eea
On the other hand, in the same manner, one gets
\be
\mathcal{L}_{\hat{\xi}} \hat{A}_{(n-1)\mu_1\cdot\cdot\cdot\mu_{(n-2)}z}
=\mathcal{L}_{\xi} A_{(n-2)\mu_1\cdot\cdot\cdot\mu_{(n-2)}}
=0 \, .\label{LiedeHatA2}
\ee

\section{The Kaluza-Klein reduction of the Lagrangian with
$p$-form potentials}\label{appendD}

In this appendix, according to the Kaluza-Klein theory, we perform a dimensional
reduction to the general Lagrangian including multiple $p$-form potentials
and scalar fields from $(D+1)$ dimensions to $D$ dimensions on
the circle $S^1$. As before, the $(D+1)$-dimensional spacetime is endowed
with the coordinate system $\hat{x}^{\hat{\mu}}=(x^\mu,z)$,
and the $(D+1)$-dimensional  metric ansatz is still
supposed to take the form given in Eq. (\ref{Dplus1metric}).

As a warm-up, we prove that the contraction between the
$(D+1)$-dimensional totally antisymmetric $n$-rank tensors
$\hat{\Phi}_{(n)}^{\hat{\mu}_1\cdot\cdot\cdot\hat{\mu}_{n}}$ and
$\hat{\Psi}^{(n)}_{\hat{\mu}_1\cdot\cdot\cdot\hat{\mu}_{n}}$ can
be expressed as
\bea
\hat{\Phi}_{(n)}^{\hat{\mu}_1\cdot\cdot\cdot\hat{\mu}_{n}}
\hat{\Psi}^{(n)}_{\hat{\mu}_1\cdot\cdot\cdot\hat{\mu}_{n}}&=&
\hat{g}^{\mu_2\nu_2}\cdot\cdot\cdot\hat{g}^{\mu_n\nu_n}
\Big(\hat{g}^{\mu_1\nu_1}\hat{\Phi}^{(n)}_{\nu_1\cdot\cdot\cdot\nu_{n}}
\hat{\Psi}^{(n)}_{\mu_1\cdot\cdot\cdot\mu_{n}}
+n\hat{g}^{zz}
\hat{\Phi}^{(n)}_{z\nu_2\cdot\cdot\cdot\nu_{n}}
\hat{\Psi}^{(n)}_{z\mu_2\cdot\cdot\cdot\mu_{n}}
\nn \\
&&+n\hat{g}^{\mu_1z}
\hat{\Phi}^{(n)}_{z\nu_2\cdot\cdot\cdot\nu_{n}}
\hat{\Psi}^{(n)}_{\mu_1\cdot\cdot\cdot\mu_{n}}
+n\hat{g}^{z\nu_1}
\hat{\Phi}^{(n)}_{\nu_1\cdot\cdot\cdot\nu_{n}}
\hat{\Psi}^{(n)}_{z\mu_2\cdot\cdot\cdot\mu_{n}}
\Big)
\nn \\
&&+\frac{n(n-1)}{2}
\hat{g}^{\mu_1z}\hat{g}^{z\nu_2}
\hat{g}^{\mu_3\nu_3}\cdot\cdot\cdot\hat{g}^{\mu_n\nu_n}
\hat{\Phi}^{(n)}_{z\nu_2\cdot\cdot\cdot\nu_{n}}
\hat{\Psi}^{(n)}_{\mu_1z\mu_3\cdot\cdot\cdot\mu_{n}}
\nn \\
&&+\frac{n(n-1)}{2}
\hat{g}^{z\nu_1}\hat{g}^{\mu_2z}
\hat{g}^{\mu_3\nu_3}\cdot\cdot\cdot\hat{g}^{\mu_n\nu_n}
\hat{\Phi}^{(n)}_{\nu_1z\nu_3\cdot\cdot\cdot\nu_{n}}
\hat{\Psi}^{(n)}_{z\mu_2\cdot\cdot\cdot\mu_{n}}
\, . \label{Dplus1contrPP}
\eea
Moreover, if
\bea
\hat{\Phi}_{(n)\mu_1\cdot\cdot\cdot\mu_{n}}&=&
\Phi_{(n)\mu_1\cdot\cdot\cdot\mu_{n}}\, , \quad
\hat{\Phi}_{(n)\mu_1\cdot\cdot\cdot\mu_{(n-1)}z}=
\Phi_{(n-1)\mu_1\cdot\cdot\cdot\mu_{(n-1)}}
\, , \nn \\
\hat{\Psi}_{(n)\mu_1\cdot\cdot\cdot\mu_{n}}&=&
\Psi_{(n)\mu_1\cdot\cdot\cdot\mu_{n}}\, , \quad
\hat{\Psi}_{(n)\mu_1\cdot\cdot\cdot\mu_{(n-1)}z}=
\Psi_{(n-1)\mu_1\cdot\cdot\cdot\mu_{(n-1)}}
\, , \label{PhiPsinDD}
\eea
where $\Phi_{(n)}$, $\Phi_{(n-1)}$, $\Psi_{(n)}$ and $\Psi_{(n-1)}$
are totally antisymmetric tensors in $D$-dimensional spacetime,
Eq. (\ref{Dplus1contrPP}) can be simplified as
\bea
\hat{\Phi}_{(n)}^{\hat{\mu}_1\cdot\cdot\cdot\hat{\mu}_{n}}
\hat{\Psi}^{(n)}_{\hat{\mu}_1\cdot\cdot\cdot\hat{\mu}_{n}}&=&
e^{-2n\alpha\varphi}
\Big(\Phi_{(n)}^{\mu_1\cdot\cdot\cdot\mu_{n}}
-n\Phi_{(n-1)}^{[\mu_1\cdot\cdot\cdot\mu_{(n-1)}}
\mathcal{A}^{\mu_{n}]}\Big)
\nn \\
&&\times\Big(\Psi^{(n)}_{\mu_1\cdot\cdot\cdot\mu_{n}}
-n\Psi^{(n-1)}_{[\mu_1\cdot\cdot\cdot\mu_{(n-1)}}
\mathcal{A}_{\mu_{n}]}\Big)
\nn \\
&&+ne^{2(D-n-1)\alpha\varphi}
\Phi_{(n-1)}^{\mu_1\cdot\cdot\cdot\mu_{(n-1)}}
\Psi^{(n-1)}_{\mu_1\cdot\cdot\cdot\mu_{(n-1)}}
\, . \label{Dplus1contrPP1}
\eea

Now we move on to consider the Kaluza-Klein reduction of the
$(D+1)$-dimensional Lagrangian for $p$-form potentials $\hat{A}_{(p)}$
and scalar fields $\phi^{(k)}$ $(k=1,\cdot\cdot\cdot,m)$, which
is postulated to take the general form
\be
\hat{\mathcal{L}}_{\hat{M}}=\sqrt{-\hat{g}}
W\big(\phi^{(k)}\big)\hat{H}_{(N)}^{\hat{\mu}_1\cdot\cdot\cdot\hat{\mu}_N}
\hat{Y}_{(N)\hat{\mu}_1\cdot\cdot\cdot\hat{\mu}_N}
\, , \label{Dplus1LagranLM}
\ee
where the two totally antisymmetric $N$-rank tensors $\hat{H}_{(N)}$ and
$\hat{Y}_{(N)}$ in $(D+1)$-dimensional spacetime are read off as
\bea
\hat{H}_{(N)\hat{\mu}_1\cdot\cdot\cdot\hat{\mu}_N}&=&
\big(\hat{F}_{(q_1)}\cdot\cdot\cdot \hat{F}_{(q_t)}
\hat{A}_{(p_1)}\cdot\cdot\cdot \hat{A}_{(p_s)}\big)_{[\hat{\mu}_1\cdot\cdot\cdot\hat{\mu}_N]}
\, , \nn \\
\hat{Y}_{(N)\hat{\mu}_1\cdot\cdot\cdot\hat{\mu}_N}&=&
\big(\hat{F}_{(\tilde{q}_1)}\cdot\cdot\cdot \hat{F}_{(\tilde{q}_j)}
\hat{A}_{(\tilde{p}_1)}\cdot\cdot\cdot
\hat{A}_{(\tilde{p}_i)}\big)_{[\hat{\mu}_1\cdot\cdot\cdot\hat{\mu}_N]}
\, , \label{Dplus1HYNform}
\eea
taking the similar structures as the ones given in Eq. (\ref{HYNform}). In
Eq. (\ref{Dplus1HYNform}), all the potentials are defined by
\bea
\hat{A}_{(p_c)}&=&A_{(p_c)}+A_{(p_c-1)}\wedge dz \, , \quad
1\leq c \leq s
\, , \nn \\
\hat{A}_{(\tilde{p}_c)}&=&A_{(\tilde{p}_c)}
+A_{(\tilde{p}_c-1)}\wedge dz \, , \quad
1\leq c \leq i
\, , \nn \\
\hat{A}_{(q_c-1)}&=&A_{(q_c-1)}+A_{(q_c-2)}\wedge dz \, , \quad
1\leq c \leq t
\, , \nn \\
\hat{A}_{(\tilde{q}_c-1)}&=&A_{(\tilde{q}_c-1)}
+A_{(\tilde{q}_c-2)}\wedge dz \, , \quad
1\leq c \leq j
\, , \label{Dplus1Al}
\eea
where all the quantities $A_{(p)}$
$(p=p_c, p_c-1,\cdot\cdot\cdot,\tilde{q}_c-2)$ are the $D$-dimensional
potentials, which only depend on the $x^\mu$ coordinates. As a consequence,
the components of the $(D+1)$-dimensional field strengths
$\hat{F}_{(q_c)}$ and $\hat{F}_{(\tilde{q}_c)}$, defined through
$\hat{F}_{(q_c)}=d\hat{A}_{(q_c)}$ and
$\hat{F}_{(\tilde{q}_c)}=d\hat{A}_{(\tilde{q}_c)}$
respectively, are expressed as
\bea
\hat{F}^{(q_c)}_{\mu_1\cdot\cdot\cdot\mu_{q_c}}&=&
F^{(q_c)}_{\mu_1\cdot\cdot\cdot\mu_{q_c}}
\, , \quad
\hat{F}^{(q_c)}_{\mu_1\cdot\cdot\cdot\mu_{(q_c-1)}z}=
F^{(q_c-1)}_{\mu_1\cdot\cdot\cdot\mu_{(q_c-1)}}
\, , \nn \\
\hat{F}^{(\tilde{q}_c)}_{\mu_1\cdot\cdot\cdot\mu_{\tilde{q}_c}}&=&
F^{(\tilde{q}_c)}_{\mu_1\cdot\cdot\cdot\mu_{\tilde{q}_c}}
\, , \quad
\hat{F}^{(\tilde{q}_c)}_{\mu_1\cdot\cdot\cdot\mu_{(\tilde{q}_c-1)}z}=
F^{(\tilde{q}_c-1)}_{\mu_1\cdot\cdot\cdot\mu_{(\tilde{q}_c-1)}}
\, . \label{Dplus1fiesF}
\eea
In the above equation, the $l$-form $F_{(l)}=dA_{(l-1)}$
$(l=q_c,q_c-1,\tilde{q}_c,\tilde{q}_c-1)$ are $D$-dimensional field
strengths. Obviously, all the field strengths are independent of the $z$
coordinate. In addition, it is supposed that all the scalar fields
$\phi^{(k)}$ do not rely on the $z$ variable, yielding that the function
$W\big(\phi^{(k)}\big)$ does not depend on the $z$ coordinate too.

As in Eq. (\ref{PhiPsinDD}), we further introduce the following
$D$-dimensional totally antisymmetric tensors, which are
\bea
H_{(N)\mu_1\cdot\cdot\cdot\mu_N}&=&
\hat{H}_{(N)\mu_1\cdot\cdot\cdot\mu_N}
\, , \quad
H_{(N-1)\mu_1\cdot\cdot\cdot\mu_{(N-1)}}=
\hat{H}_{(N)\mu_1\cdot\cdot\cdot\mu_{(N-1)}z}
\, , \nn \\
Y_{(N)\mu_1\cdot\cdot\cdot\mu_N}&=&
\hat{Y}_{(N)\mu_1\cdot\cdot\cdot\mu_N}
\, , \quad
Y_{(N-1)\mu_1\cdot\cdot\cdot\mu_{(N-1)}}=
\hat{Y}_{(N)\mu_1\cdot\cdot\cdot\mu_{(N-1)}z}
\, , \label{HYinDdspac}
\eea
respectively. On basis of the $(D+1)$-dimensional metric ansatz
(\ref{Dplus1metric}), by performing Kaluza-Klein reduction on a circle,
together with help of Eq. (\ref{Dplus1contrPP1}), the higher-dimensional
Lagrangian $\hat{\mathcal{L}}_{\hat{M}}$ is rewritten in terms of the
lower-dimensional fields as
\bea
\hat{\mathcal{L}}_{\hat{M}}&=&\tilde{\mathcal{L}}_{M}
\nn \\
&=&\sqrt{-g}W\Big[e^{-2(N-1)\alpha\varphi}
\Big(H_{(N)}^{\mu_1\cdot\cdot\cdot\mu_{N}}
-NH_{(N-1)}^{[\mu_1\cdot\cdot\cdot\mu_{(N-1)}}
\mathcal{A}^{\mu_{N}]}\Big)
\nn \\
&&\times\Big(Y^{(N)}_{\mu_1\cdot\cdot\cdot\mu_{N}}
-NY^{(N-1)}_{[\mu_1\cdot\cdot\cdot\mu_{(N-1)}}
\mathcal{A}_{\mu_{N}]}\Big)
\nn \\
&&+Ne^{2(D-N)\alpha\varphi}
H_{(N-1)}^{\mu_1\cdot\cdot\cdot\mu_{(N-1)}}
Y^{(N-1)}_{\mu_1\cdot\cdot\cdot\mu_{(N-1)}}
\Big]
\, . \label{DdLagranLM}
\eea

\section{The $\hat{K}_{(\hat{F})}^{\mu\nu}$ potential
in terms of the D-dimensional fields}\label{appendE}

To get the exact relationship of the off-shell ADT potentials in
higher and lower dimensions, in this appendix, detailed calculations
are presented to derive the expression for the $(\mu,\nu)$-component
of the $(D+1)$-dimensional off-shell Noether potential
$\hat{K}_{(\hat{F})}^{\hat{\mu}\hat{\nu}}\big(\hat{\xi}_{(b)}\big)$,
defined through Eq. (\ref{Dplus1SurfKF}), in terms of all the
$D$-dimensional fields, where the
$\hat{\xi}^\mu_{(b)}$ vector, given in Eq. (\ref{Definhatximb}), has
been proved to be the $(D+1)$-dimensional Killing vector.

For convenience, four tensors $\hat{P}^{\mu\nu}_{(i)}$ $(i=1,2,3)$
and $\hat{P}^{\mu\nu}_{(b)}$ are introduced to express the
$(\mu,\nu)$-component of the $(D+1)$-dimensional off-shell
Noether potential $\hat{K}_{(\hat{F})}^{\hat{\mu}\hat{\nu}}$,
appearing in Eq. (\ref{Dplus1SurfKF}), as the following form
\be
\hat{K}_{(\hat{F})}^{\mu\nu}\big(\hat{\xi}_{(b)}\big)=
-2n(n-1)\gamma\big[\big(\hat{P}^{\mu\nu}_{(1)}
+\hat{P}^{\mu\nu}_{(2)}
+\hat{P}^{\mu\nu}_{(3)}\big)
+b\hat{P}^{\mu\nu}_{(b)}\big]
\, , \label{HatKFmunu1}
\ee
where the quantities $\hat{P}^{\mu\nu}_{(i)}$ $(i=1,2,3)$ are
given by
\bea
\hat{P}^{\mu\nu}_{(1)}&=&
\hat{g}^{\mu\alpha}\hat{g}^{\nu\beta}
\hat{F}_{(n)\alpha\beta}^{~~~~~\hat{\mu}_1\cdot\cdot\cdot\hat{\mu}_{(n-2)}}
\xi^\sigma\hat{A}_{(n-1)\sigma\hat{\mu}_1\cdot\cdot\cdot\hat{\mu}_{(n-2)}}
\, , \nn \\
\hat{P}^{\mu\nu}_{(2)}&=&
\hat{g}^{\mu\alpha}\hat{g}^{\nu z}
\hat{F}_{(n)\alpha z}^{~~~~~\hat{\mu}_1\cdot\cdot\cdot\hat{\mu}_{(n-2)}}
\xi^\sigma\hat{A}_{(n-1)\sigma\hat{\mu}_1\cdot\cdot\cdot\hat{\mu}_{(n-2)}}
\, , \nn \\
\hat{P}^{\mu\nu}_{(3)}&=&
\hat{g}^{\mu z}\hat{g}^{\nu\beta}
\hat{F}_{(n)z\beta}^{~~~~~\hat{\mu}_1\cdot\cdot\cdot\hat{\mu}_{(n-2)}}
\xi^\sigma\hat{A}_{(n-1)\sigma\hat{\mu}_1\cdot\cdot\cdot\hat{\mu}_{(n-2)}}
\, ,
\eea
and the quantity $\hat{P}^{\mu\nu}_{(b)}$ is read off as
\bea
\hat{P}^{\mu\nu}_{(b)}&=&
\Big(\hat{g}^{\mu\alpha}\hat{g}^{\nu\beta}
\hat{F}_{(n)\alpha\beta}^{~~~~~\hat{\mu}_1\cdot\cdot\cdot\hat{\mu}_{(n-2)}}
+\hat{g}^{\mu\alpha}\hat{g}^{\nu z}
\hat{F}_{(n)\alpha z}^{~~~~~\hat{\mu}_1\cdot\cdot\cdot\hat{\mu}_{(n-2)}}
\nn \\
&&+\hat{g}^{\mu z}\hat{g}^{\nu\beta}
\hat{F}_{(n)z\beta}^{~~~~~\hat{\mu}_1\cdot\cdot\cdot\hat{\mu}_{(n-2)}}
\Big)
\hat{A}_{(n-1)z\hat{\mu}_1\cdot\cdot\cdot\hat{\mu}_{(n-2)}}
\nn \\
&=&(-1)^n e^{-2n\alpha\varphi}
\tilde{F}_{(n)}^{\mu\nu\mu_1\cdot\cdot\cdot\mu_{(n-2)}}
A^{(n-2)}_{\mu_1\cdot\cdot\cdot\mu_{(n-2)}}
\, . \label{PbmunuinhatKF}
\eea
To get the last equality in the above equation, Eq. (\ref{Dplus1contrPP})
has been used. By utilizing Eq. (\ref{Dplus1contrPP}) once again, we
obtain that
\bea
\hat{P}^{\mu\nu}_{(1)}&=&
e^{-2n\alpha\varphi}\xi^\sigma \Big[
A^{(n-1)}_{\sigma\mu_1\cdot\cdot\cdot\mu_{(n-2)}}
\Big(F_{(n)}^{\mu\nu\mu_1\cdot\cdot\cdot\mu_{(n-2)}}
-(n-2)F_{(n-1)}^{\mu\nu\mu_1\cdot\cdot\cdot\mu_{(n-3)}}
\mathcal{A}^{\mu_{(n-2)}}
\Big)
\nn \\
&&-(n-2)A^{(n-2)}_{\sigma\mu_1\cdot\cdot\cdot\mu_{(n-3)}}
\Big((n-3)
F_{(n-1)}^{\mu\nu\mu_1\cdot\cdot\cdot\mu_{(n-4)}\rho}
\mathcal{A}_\rho\mathcal{A}^{\mu_{(n-3)}}
\nn \\
&&+F_{(n)}^{\mu\nu\mu_1\cdot\cdot\cdot\mu_{(n-3)}\rho}
\mathcal{A}_\rho
-\mathcal{A}^2
F_{(n-1)}^{\mu\nu\mu_1\cdot\cdot\cdot\mu_{(n-3)}}
\Big)\Big]
\nn \\
&&+(n-2)e^{2(D-n-1)\alpha\varphi}
F_{(n-1)}^{\mu\nu\mu_1\cdot\cdot\cdot\mu_{(n-3)}}
\xi^\sigma A^{(n-2)}_{\sigma\mu_1\cdot\cdot\cdot\mu_{(n-3)}}
\, , \label{ValueofP1mn}
\eea
and the quantities $\hat{P}^{\mu\nu}_{(2)}$ and $\hat{P}^{\mu\nu}_{(3)}$
take the following values
\bea
\hat{P}^{\mu\nu}_{(2)}&=&
-e^{-2n\alpha\varphi}\xi^\sigma \mathcal{A}^\nu
F_{(n-1)}^{\mu\mu_1\cdot\cdot\cdot\mu_{(n-2)}}\Big(
A^{(n-1)}_{\mu_1\cdot\cdot\cdot\mu_{(n-2)}\sigma}
+(n-2)A^{(n-2)}_{\mu_1\cdot\cdot\cdot\mu_{(n-3)}\sigma}
\mathcal{A}_{\mu_{(n-2)}}
\Big)
\, , \nn \\
\hat{P}^{\mu\nu}_{(3)}&=&
-\hat{P}^{\mu\nu}_{(2)}(\mu\leftrightarrow\nu)
\, .\label{ValueofP23mn}
\eea

Through a combination of Eqs. (\ref{PbmunuinhatKF}), (\ref{ValueofP1mn})
and (\ref{ValueofP23mn}), the off-shell Noether potential
$\hat{K}_{(\hat{F})}^{\mu\nu}\big(\hat{\xi}_{(b)}\big)$ can be rewritten
in terms of the $D$-dimensional fields as the form
\be
\hat{K}_{(\hat{F})}^{\mu\nu}\big(\hat{\xi}_{(b)}\big)=
e^{-2\alpha\varphi}K_{(F)}^{\mu\nu}\big(\xi\big)
-2bn(n-1)
\gamma e^{-2n\alpha\varphi}
\tilde{F}_{(n)}^{\mu\mu_1\cdot\cdot\cdot\mu_{(n-2)}\nu}
A^{(n-2)}_{\mu_1\cdot\cdot\cdot\mu_{(n-2)}}
\, . \label{HatKFintermKF}
\ee
Here the $n$-form field strength $\tilde{F}_{(n)}$ is given by
Eq. (\ref{TildeFnform}).


\end{document}